\documentclass[a4paper,12pt]{article}
\usepackage[margin=1in]{geometry}
\usepackage{titlesec}
\usepackage{lipsum}
\usepackage{csquotes}

\usepackage{amssymb,amsmath,mathrsfs}                                                               
\usepackage[title]{appendix}
\usepackage{xcolor}                                            
\usepackage{graphicx}
        
\makeatletter
\renewcommand\@biblabel[1]{}
\makeatother
          
\usepackage{subcaption}
\usepackage{multirow}
   \usepackage[affil-it]{authblk}

\usepackage{lineno}
\usepackage[final,notref,notcite]{showkeys}
\usepackage{wasysym}
\usepackage{natbib}
  
\usepackage{longtable,amssymb,graphicx,multirow,eurosym}
\usepackage{graphicx}   
\usepackage{float}   
\usepackage{appendix}      
 
\usepackage{longtable,amssymb,graphicx,multirow,eurosym}
\usepackage{graphicx}   
\usepackage{float}   
\usepackage{appendix}
\usepackage{float}     
\usepackage[multiple]{footmisc}
\usepackage{epigraph}
\usepackage{titlesec}
\titleformat{\subsection}[hang]{\itshape\normalfont\normalsize}{\thesubsection}{1em}{}

\usepackage{booktabs}

\usepackage{tikz-cd}
\usepackage{tikz}
\usepackage{hyperref}
\usepackage{lscape}

\makeatletter
\renewcommand\@biblabel[1]{}
\makeatother
\addcontentsline{toc}{section}{Acknowledgement}
\begin{document}
\title{Generative Agents and Expectations: \\ Do LLMs Align with Heterogeneous Agent Models?}

\author{Filippo Gusella}

\affil{Università Cattolica del Sacro Cuore, Milano\\
       \smallskip Università degli Studi di Firenze\\
       \smallskip New York University Florence \\
       }

\author{Eugenio Vicario}
\affil{IMT School for Advanced Studies Lucca}

\date{}

\maketitle
	
\small

\begin{abstract}
Results in the Heterogeneous Agent Model (HAM) literature determine the proportion of fundamentalists and trend followers in the financial market. This proportion varies according to the periods analyzed. In this paper, we use a large language model (LLM) to construct a generative agent (GA) that determines the probability of adopting one of the two strategies based on current information. The probabilities of strategy adoption are compared with those in the HAM literature for the S\&P 500 index between 1990 and 2020.  Our findings suggest that the resulting artificial intelligence (AI) expectations align with those reported in the HAM literature.  At the same time, extending the analysis to artificial market data helps us to filter the decision-making process of the AI agent. In the artificial market, results confirm the heterogeneity in expectations but reveal systematic asymmetry toward the fundamentalist behavior. 
\end{abstract}   
\bigskip
\bigskip
\bigskip
\bigskip

\bigskip
\bigskip

\emph{Jel codes}: E30, E70, D84	
\bigskip

\newpage

\epigraph{
\textit{“As economics acquires aspirations to explain behavior under these typical conditions of modern organizational and public life, it will have to devote major energy to building a theory of procedural rationality to complement existing theories of substantive rationality. Some elements of such a theory can be borrowed from the neighboring disciplines of operations research, artificial intelligence, and cognitive psychology”}
}{Herbert A. Simon, \textit{Rationality as Process and as Product of Thought}, 1978}

\section{Introduction}

The role of expectations among economic agents is fundamental in economics. The ways in which agents form their beliefs not only raise fundamental questions about the observed dynamics of both real and financial variables, but also affect the role that monetary and fiscal policy is expected to play. 

For several decades, the prevailing approach to analyzing the formation of expectations has been based on the foundational concept of a representative rational agent. On the one hand, the use of the representative agent excluded any consideration of heterogeneity. On the other hand, rational expectations ruled out any endogenous influence on the dynamics of macroeconomic and financial variables \citep{muth1961rational,lucas1972expectations,fama1991efficient}. Parallel to this view, an alternative perspective on the role of expectations emerged, forming the foundation of what is known as the behavioral approach \citep{simon1978rationality}. Within this field, models of heterogeneous agents (HAM) deviate from the principle of methodological rational individualism. These models incorporate the concept of heterogeneity in expectations, with different groups of agents forming their beliefs based on heuristics \citep{day1990bulls,lux1995herd,brock1998heterogeneous}.\footnote{For a comprehensive overview of this literature, see the seminal survey by \cite{hommes2006heterogeneous}, as well as the more recent review by \cite{dieci2018heterogeneous} and \cite{hommes2021behavioral}, which incorporate recent advances in heterogeneous agent modeling.}

In the early 21st century, various estimation methods were employed to test the hypothesis of heterogeneity in expectations \citep{lux2018empirical,ellen2018heterogeneous}. These studies support theoretical models by identifying distinct beliefs among groups of speculators \citep{lux2018estimation,majewski2020co,schmitt2021heterogeneous,ter2021comparing,gusella2024endogenous,di2025sentiment}. In addition, fluctuations in asset prices can derive from a switching mechanism that moves agents among different expectation-based rules; the presence and time-varying adoption of these rules lead to the endogenous emergence of boom and bust periods \citep{schmitt2021trend,gatti2025endogenous}. In particular, while periods before the dot-com and global financial crises are associated with a high percentage of agents adopting a trend-following strategy, the subsequent post-crisis phases are instead characterized by a rising share of fundamentalists, which drives the dynamics toward a convergence of prices to their fundamental values \citep{hommes2017booms}.

In recent years, a rapidly expanding body of literature has examined the interactions between humans and artificial intelligence (AI), exploring how AI can be leveraged across diverse domains — from education and healthcare to politics and economics \citep{tsvetkova2024new,motoki2025assessing,di2025individuals}. This line of research highlights the transformative potential of AI systems as tools capable of augmenting human decision-making, generating knowledge, and shaping social and economic outcomes. On the economic side, the rapid diffusion of large language models has opened new avenues for research on expectation formation, decision-making, and forecasting, as these models can act as artificial economic agents that generate predictions, narratives, and trading strategies. Consequently, a growing number of studies have investigated whether LLMs can serve as substitutes or complements to human forecasters, evaluating their performance against professional forecasts, survey data, or controlled laboratory experiments \citep{bybee2023ghost,lopez2023can,faria2024artificial,hansen2024simulating,zarifhonarvar2024experimental,del2025can,lopez2025can}. The present study aligns with this strand of research and advances the analysis of belief formation by introducing a generative agent driven by a Large Language Model (LLM), which serves as an active decision maker in a simulated financial environment.  However, while previous research has provided valuable benchmarks on the accuracy and consistency of LLM-based forecasts, much less attention has been devoted to the behavioral mechanisms underlying the strategies adopted by LLM agents. Given the ability of LLMs to mimic human behavior and replicate cognitive biases \citep{binz2023using,strachan2024testing}, it becomes crucial to analyze their decision processes, particularly as they play an increasingly central role in economic and financial decision-making \citep{gao2024large,korinek2023generative,horton2023large}. Their ability to replicate or diverge from the behavior of past economic agents may either reinforce existing market dynamics or counteract them. Indeed, LLMs may default to the most statistically dominant interpretation, often aligning with the prevailing mainstream or dominant view \citep{shen2024understanding,qu2024performance}.
This paper contributes to this growing literature by examining whether generative agent price strategies align with those documented in the heterogeneous agent modeling literature, thereby providing new insights into LLM-driven market behavior.

We proceed as follows. In a simulated financial market environment, the generative LLM agent can select between two market strategies: one assumes convergence between future prices and perceived fundamental values. This is the expectation of \emph{fundamentalists}.  The other, associated with the behavior of \emph{trend followers}, follows a trend-based approach, where speculation is based on past price movements.\footnote{The focus on these two types of expectations is motivated by the fact that most empirical studies on heterogeneous agent models adopt the classification of trend followers and fundamentalists, as established in the theoretical literature.} Choosing between these strategies, we examine how the LLM agent forms beliefs when forecasting S\&P 500 Index over the period January 1990 to December 2020 and compare its decision with traditional econometric results in the literature on the heterogeneous agent model. In doing this, we offer opportunities to test theories and hypotheses on expectation formation, helping to determine whether LLM-driven agents align with or diverge from established HAM results, which may therefore have been inherited in ChatGPT training.  At the same time, based on artificial market conditions, we aim to deepen the analysis by examining whether it is possible to identify the behavioral rules that govern the adoption of specific expectations based on available information. This helps us to understand the specific mechanisms that underlie the AI agent's decision-making process. In other words, the analysis allowed us to filter the behavioral rules that have been internalized by the generative agent from heterogeneous agent models. 

Results based on time series analysis indicate that AI-generated expectations align with findings from the HAM literature. In particular, we show that LLM-based expectations mirror key deviations from the fundamentalist expectation during major events such as the dot-com bubble and the global financial crisis. Furthermore, when the expectations of the generative agents are explicitly modeled, the values of the reaction parameters are found to be in line with those predicted in the HAMs setting. At the same time, results based on artificial data confirm the heterogeneity in expectations, but with an asymmetry toward fundamentalist behavior.   

The remainder of the paper is organized as follows. Section 2 introduces the beliefs formalized in the HAM literature, reviews related empirical findings, and outlines the AI setting. Section 3 presents the simulation results. Section 4 concludes.
\newpage

\section{Methodology}

\subsection{\emph{Theoretical setting and empirical results }}

In the literature on heterogeneous agent models, developed through early contributions by \cite{day1990bulls,chiarella1992dynamics,lux1995herd,brock1998heterogeneous}, the asset price dynamics is shaped by the coexistence of groups of agents with different expectations. The most common specification distinguishes between two of them: fundamentalist ($F$) and trend follower ($TF$) expectations.\footnote{Although related to the broader Agent-Based Modeling (ABM) literature \citep{dawid2018agent,dosi2019more,delli2018agent,monti2023learning}, HAMs do not consider numerous interacting agents, but rely on a smaller set of behavioral rules adopted by groups of agents. This formalization allows analytical insights into how different expectations shape macro-finance outcomes.}  Fundamentalists rely on the efficient market hypothesis, assuming that the price ($P$) will eventually converge to its fundamental value ($P_t^F$). This expectation can be formalized as follows:

\begin{equation}\label{eq: alpha}
   {E^F}\left[ {{P_{t + 1}}} \right] = {P_t} + {\alpha _t}\left( {P_t^F - {P_t}} \right),\quad \quad 0 \le \alpha  \le 1,
\end{equation}
where $\alpha$, the reaction parameter, captures the intensity with which the market price is driven toward the fundamental value. This implies that during periods of pronounced asset price expansions or contractions, fundamentalist agents anticipate a correction, expecting the price to realign with the underlying fundamental.

In contrast, trend followers base their forecasts on previously observed price patterns. They extrapolate recent trends and assume that increases (or decreases) in past prices are likely to persist in the near future: \begin{equation}\label{eq: beta}
    E^{TF}\left[ P_{t+1}\right]=P_t+ \beta_t \left( P_t-P_{t-1} \right),\quad \quad \beta>0.
\end{equation}

The strength of this behavior is governed by the reaction parameter $\beta$, which captures how aggressively trend followers respond to past price movements.

Aggregating these two behavioral rules, the market price at time $t+1$ can be expressed as a weighted average of the expectations of the two groups:
\begin{equation}\label{eq: price_exp}
{P_{t+1}} = {\delta _t}{E^F}\left( {{P_{t + 1}}} \right) + \left( {1 + {\delta _t}} \right){E^{TF}}\left( {{P_{t + 1}}} \right),
\end{equation}
where $\delta _t$ and $1-\delta _t$ are the percentage of fundamentalists and trend followers in the market, respectively. Combining Eqs. \ref{eq: alpha}, \ref{eq: beta}, and \ref{eq: price_exp} yield the following reduced-form expression for price dynamics:

\begin{equation}\label{eq: price}
{P_{t + 1}} = {P_t} + {\delta _t}\alpha_t\left( {P_t^F - {P_t}} \right) + \left( {1 + {\delta _t}} \right)\beta_t\left( {{P_t} - {P_{t - 1}}} \right).
\end{equation}

The coexistence of heterogeneous behavior, captured by differing reaction parameters and a changing distribution of agents over time, gives rise to nonlinear market dynamics that produce market booms and crashes. 

Building on this theoretical framework, over the last three decades, a growing body of empirical work has investigated whether the HA models are consistent with observed asset price dynamics.\footnote{The empirical performance of HAMs in equity markets has stimulated a wider range of applications across different markets, including commodities \citep{reitz2007commodity}, foreign exchange \citep{goldbaum2014empirical}, housing market \cite{bolt2014identifying}, and macro model \citep{cornea2019behavioral,kukacka2023estimation,lux2024lack}.} These contributions generally adopt the two forecasting strategies previously defined to keep the model tractable and to prevent over-parameterization. Essentially, applying equation \ref{eq: price} to the data reveals two main findings: first, investors hold heterogeneous beliefs rather than a single representative expectation.  Empirical findings generally indicate that fundamentalist behavior stabilizes asset prices, whereas chartist behavior destabilizes them, causing prices to deviate from the fundamental value. This is reflected in significant, theoretically consistent estimates for both types of reaction parameters. In particular, the reaction coefficient for fundamentalists falls within the expected range of 0 to 1 \citep{franke2011estimation,ter2021comparing,kukacka2017estimation}. The exact value varies depending on the specific asset considered. Conversely, the reaction parameter for trend followers is typically positive  \citep{chiarella2014heterogeneous,recchioni2015calibration,gusella2024endogenous}. However, when this parameter is not subject to certain restrictions, it can also have negative values, allowing contrarian behavior \citep{chiarella2012estimating}. Second, agents frequently switch between fundamentalist and trend-follower rules. The prevailing behavior depends on the market environment and the period under analysis. Studies focusing on the late 1990s and early 2000s show that episodes such as the dot-com bubble and the global financial crisis were preceded by a surge in trend-following behavior, which contributed to the formation of asset price bubbles. However, after such crises, the price strategy tended to revert toward fundamentalist expectations, leading to more stable price adjustments \citep{boswijk2007behavioral,chiarella2014heterogeneous,hommes2017booms,gatti2025endogenous}.

\subsection{\emph{AI setting}}

As previously outlined, this paper aims to evaluate whether, and to what extent, the results and assumptions found in the heterogeneous agent modeling  literature have been internalized within the emergent knowledge of major commercial large language models. The primary focus of our analysis is the composition of the market: specifically, whether the LLM chooses a relative weighting between fundamentalists and trend followers that aligns with established findings in the literature.
In addition, we extract the expectations conditional on the strategy employed. From these expectations, we derive the reaction parameters of the equations that, in the HAM literature, characterize the behavior of fundamentalists and trend followers. Combining the  market composition with these reaction parameters enables us to compute the expected price for the following period, under the hypothetical scenario of a market populated exclusively by generative agents relying on an LLM to make their trading decisions.

The simulation analysis employs the ChatGPT-4o-mini model, accessed via an API key through a dedicated Python script. Interaction with the model is achieved through the use of prompting techniques. Two distinct prompts are used to maintain analytical clarity and isolate the mechanisms under investigation.
The first prompt is designed to elicit the market composition, that is, the relative shares of fundamentalists and trend followers within the market. Specifically, we input the following prompt:\\

\noindent
\textbf{Prompt 1 — Market Composition:} \textit{You are a financial trader and you need to choose between two alternative market strategies according to your analysis based on expectation formation: the first is a fundamentalist approach, and the second is a trend-following approach. You have access to the current value of the fundamental, the current value of the asset price, and the value of the asset price at the previous period. In conclusion, for each strategy, you only have to provide a weight that represents the probability with which you would choose the strategy. The sum of the weights must be 1. Do not provide any other comment, just the two weights divided by a comma.} \\

Following the prompt, three values are provided: the current price, the price of the previous period, and the current fundamental value. The price is the value of S\&P 500 index at monthly frequency from 1990 to 2020, while the fundamental price is calculated through the Gordon growth model as in Chiarella et al. (2012) using S\&P 500 data on dividends. Each interaction includes only these values, with no memory of the previous input. This prevents identifying the overall market context, extrapolating the market phase, or linking responses to prior literature on market structure.

The second prompt is used to obtain the model's expectations of the future price, conditional on the two distinct market strategies. As before, the prompt is followed by the two price values and the value of the fundamental: \\

\noindent
\textbf{Prompt 2 — Price Expectations:} \textit{You are a financial trader and you need to provide the value of price expectations for the next period using two alternative market strategies according to your analysis: the first is a fundamentalist approach, and the second is a trend-following approach. You have access to the current value of the fundamental, the current value of the asset price, and the value of the asset price at the previous period. In conclusion, you only have to provide the value of price expectations for each strategy. Do not provide any other comment or word, just the two numbers divided by a comma.} \\

In the existing literature, the composition of the market and the reaction parameters are typically estimated simultaneously \citep{lux2018empirical,ellen2018heterogeneous}. In contrast, this study deliberately employs separate prompts, ensuring that the two dimensions of the analysis remain independent. This methodological choice is motivated by our primary interest in market composition, which represents a more straightforward and interpretable concept from which implicit heuristics and behavioral rules can be more easily inferred. 
Conversely, deriving reaction parameters from expectations involves additional analytical complexity, as it requires the specification of functional forms for the reaction equations, introducing further uncertainty into the interpretation of results. By adopting separate prompts, we ensure that the outcomes concerning market composition are not confounded by ancillary information provided to or requested from the LLM. 
For completeness, in the Appendix \ref{appendix: prompt} we also report the results obtained using a single integrated prompt, which simultaneously generates both market composition and expectations. Although minor quantitative differences arise, the main qualitative findings remain robust.

To obtain the results, we employed the ChatGPT \textit{ gpt-4o-mini} model. Robustness checks using an alternative model are reported in the Appendix \ref{appendix: appendix 3.5}. The temperature parameter was set to 1, corresponding to the default configuration. In models such as ChatGPT, the temperature parameter regulates the degree of randomness in text generation by influencing the likelihood of selecting less probable tokens. At low values, close to zero, the model behaves deterministically, producing stable, predictable, and coherent outputs, though at the expense of creativity and diversity. At higher values, typically between 0.8 and 1.2, the model explores a broader range of possible continuations, increasing variability and originality in the responses, but also the likelihood of inconsistencies or errors. In the Appendix \ref{appendix: appendix temperature}, we test the robustness of our results with respect to different temperature settings. 

The results were obtained through 50 independent replications of the prompt for each value triplet and the average of the generated responses was used as the final result. In the Appendix \ref{appendix: dimension}, we show that 50 replications are sufficient to stabilize the results. 
Finally, consistent with the discussion of the integrated prompt presented earlier, in the appendix \ref{appendix: prompt} we also report robustness analyzes that assess the sensitivity of the results to variations in the prompt formulation.

\section{Results}

In this section, we present the results of the simulations, which are divided into two parts. In subsection \ref{section: s&p}, we apply the proposed procedure to the time series of the S\&P 500 Index and the corresponding Gordon fundamental value over the period 1990 to 2020. In this way, we identify the market composition across different time periods (Prompt 1). We then extract the reaction parameters of the agents using the expected price for the following period (Prompt 2), which allows us to analyze the forecasting error of the generative agent. Through the use of Prompt 1, we obtain a qualitative insight into the relative presence of fundamentalists and trend followers, whereas Prompt 2 provides a quantitative result. In subsection \ref{section: artificial market}, we investigate the implicit heuristics and behavioral rules employed by the LLM. To this end, we generate artificial market conditions and examine how the model responses vary as these conditions change. As before, we begin by simulating the market composition (Prompt 1) and then, using the agents’ expectations (Prompt 2), we derive the reaction parameters and compute the expected price variation for the subsequent period.

\subsection{Standard \& Poor 500 index}\label{section: s&p}

\begin{figure}[H]
\centering
\includegraphics[scale=0.4]{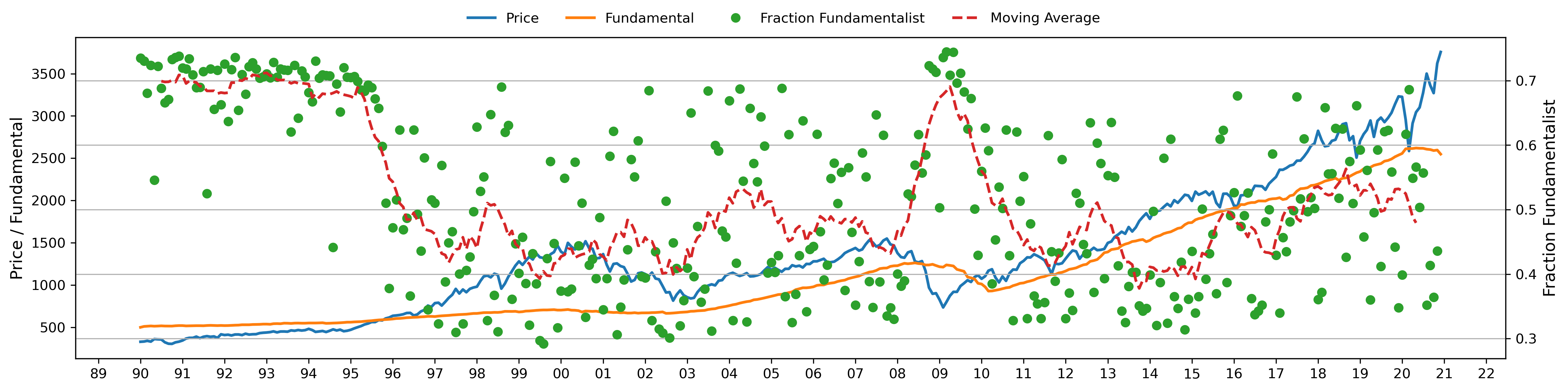}
\caption{Time series of the S\&P 500 index and the Gordon fundamental value, together with the percentage of adoption of the fundamentalist expectation.}\label{fig: percentage}
\end{figure}

Figure \ref{fig: percentage} plots the dynamics of the S\&P 500 Index (blue line) together with the corresponding Gordon Fundamental Value (orange line). Together with these two dynamics, the green dots represent the probability with which the generative agent selects the fundamentalist strategy when forecasting the next period's price, based on the information provided through the data prompt. These probabilities are computed as the average result of 50 Monte Carlo simulations. The red dashed line reports the 13-month centered moving average of this probability measure, calculated over a rolling window. 

Several interesting dynamics emerge over the sample period. Focusing on the evolution of expectation formation, the figure shows that during the first half of the 1990s, the probability of observing a  generative fundamentalist behavior is relatively high, stabilizing around a value of 0.7, which is consistent with prices moving closely toward the fundamental value. However, in the second half of the same decade, there is a marked increase in the likelihood of adopting a generative trend-following expectation. This trend-following dominance persists through the late 1990s, with peaks just before the dot-com crisis, suggesting that boom dynamics were fueled by agents extrapolating from rising prices. Furthermore, this same behavior appears to dominate in the immediate aftermath, indicating that the bust phase was largely driven by bearish trend-following expectations. 
A similar pattern emerges in the run-up to the global financial crisis, when generative trend-following expectations again prevail. In particular, although the probability of fundamentalist behavior occasionally rises thereafter, generative trend-following strategies remain relatively strong in the years preceding the global financial crisis. In contrast, the 2007-2008 collapse is followed by a more persistent reversion to fundamentalist behavior, restoring convergence between prices and fundamentals.

Overall, the strategies adopted by the LLM agent mirror patterns often reported in the empirical literature on heterogeneous agent models. In particular, the evidence suggests that boom phases are fueled by trend-following behavior that drives prices away from fundamentals, as seen before the dot-com bubble and the global financial crisis. Conversely, fundamentalist behavior tends to dominate after the collapse of speculative booms, restoring the link between prices and underlying values \citep{boswijk2007behavioral,chiarella2014heterogeneous,hommes2017booms,gatti2025endogenous}. It is worth noting that this last mechanism is more clearly reflected in the aftermath of the global financial crisis than in the post-dot-com period, highlighting asymmetries across different market events.

As a further analysis, we apply the Prompt 2, related to price expectations, to the historical series of the S\&P 500 Index and its corresponding fundamental value. From the expectations obtained through 50 Monte Carlo simulations, we computed the reaction parameters, as previously defined in Equations \ref{eq: alpha} and \ref{eq: beta}, such that:

\begin{equation}
    \alpha_t = \frac{E^F[P_{t+1}]-P_t}{P^f_t-P_t} 
\end{equation}
and
\begin{equation}
    \beta_t = \frac{E^{TF}[P_{t+1}]-P_t}{P_t-P_{t-1}} 
\end{equation} 

\begin{figure}[H]
\centering
\includegraphics[scale=0.4]{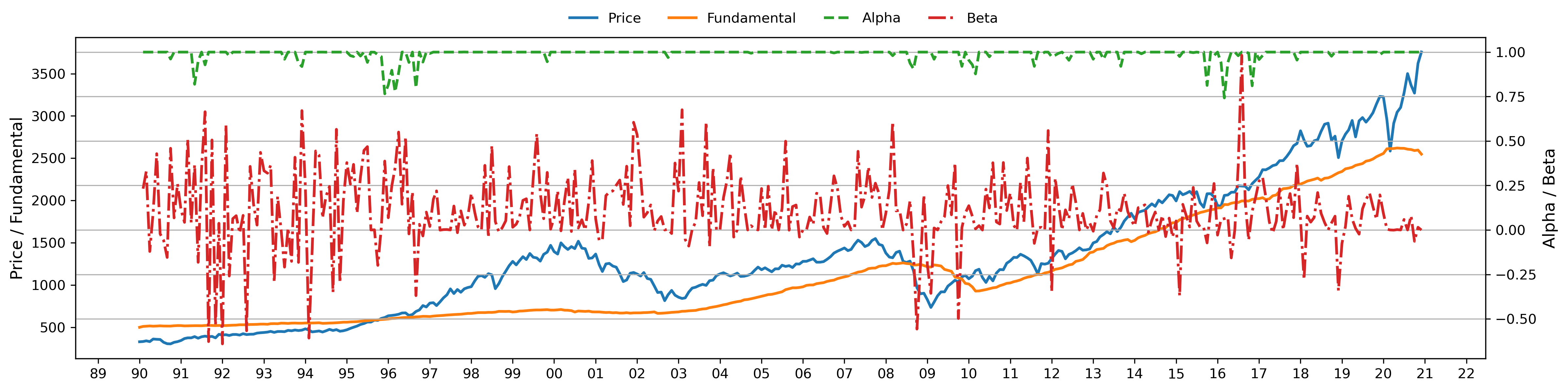}
\includegraphics[scale=0.4]{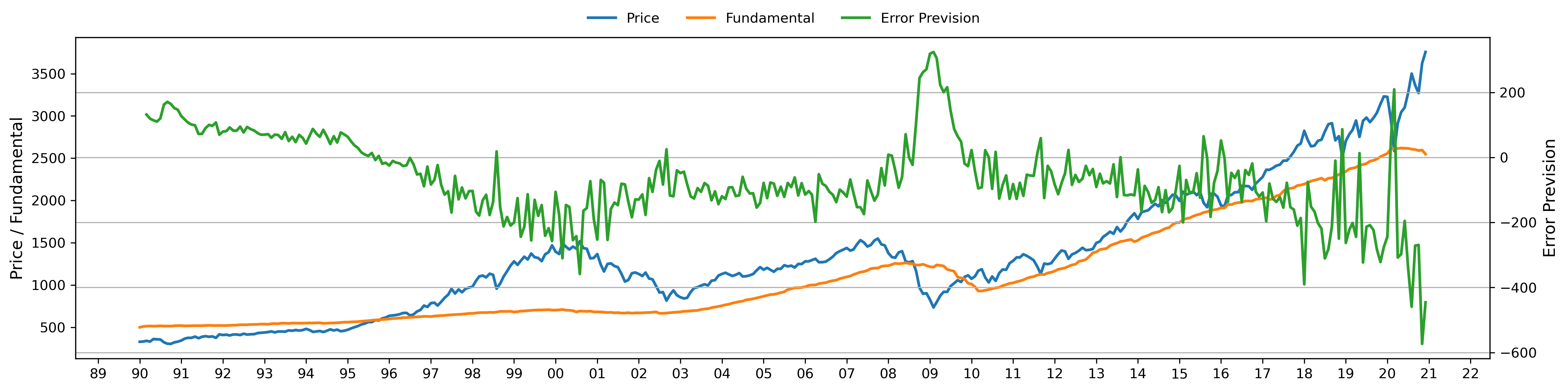}
\caption{Top: alpha and beta; Bottom: Error prevision.}\label{fig: error prevision}
\end{figure}

The estimated reaction parameters are presented in Figure \ref{fig: error prevision}. The values obtained fall within the ranges estimated in the heterogeneous agent modeling literature \citep{franke2011estimation,ter2021comparing,kukacka2017estimation}. In particular, the reaction parameter for fundamentalists lies between zero and one, with values very close to one, thereby formalizing the behavior typically associated with so-called dogmatic fundamentalists. In contrast, the estimated parameters for chartists take both positive and negative values, thus allowing for the possibility of contrarian behavior \citep{chiarella2012estimating}.
Subsequently, using the relative weight of fundamentalists together with the reaction parameters, we calculate the expected price for the following period, under the assumption that the market is populated by generative agents behaving according to the patterns observed in the simulations. By subtracting the realized price from the expected price, we derived the forecast error of the generative agents, as shown in Figure \ref{fig: error prevision}.

We observe that the sign of the forecasting errors systematically depends on the sign of the deviation between the fundamental value and the current price: when the market price lies above the fundamental, the forecasting error is negative, whereas it becomes positive when the fundamental value exceeds the market price. This pattern is driven by the consistent presence of fundamentalist agents characterized by reaction parameters close to one, that is, strict fundamentalists. These agents systematically push the expected price toward the fundamental, thereby generating a systematic bias in the forecast.

\subsection{Artificial market conditions}\label{section: artificial market}
\begin{figure}[H]
\centering
\includegraphics[scale=0.5]{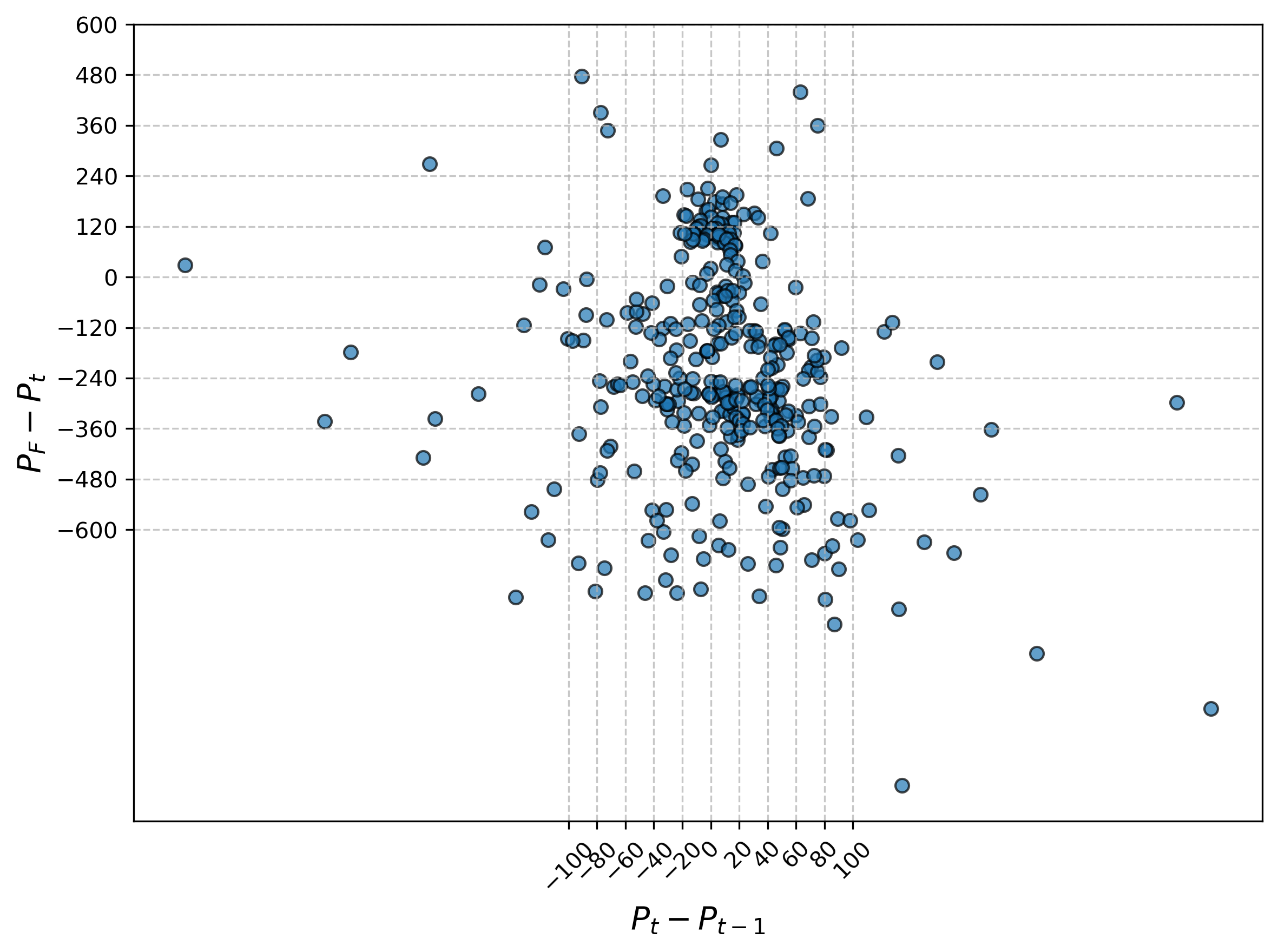}
\caption{Scatter plot of S\&P 500 price variations relative to Gordon fundamentals and lagged prices over the period 1990-2020.}\label{fig: scatter}
\end{figure}

Building on the previous  results, we deepen the analysis by investigating whether it is possible to identify the rules governing the adoption of specific expectations based on the information available to the generative agent. This approach allows for a closer examination of the mechanisms underlying the generative expectation formation, with particular attention to the conditions that trigger the switching between fundamentalist and trend-following behavior.
To uncover the implicit switching rules and heuristics driving these outcomes, we test the AI’s behavior and corresponding price strategies by varying the fundamental value and the previous price while keeping the current price constant. Specifically, we evaluate all combinations in which the distance between the fundamental value and the current price ranges from –600 to 600, with increments of 120, while the difference between the current and previous prices ranges from –100 to 100, with increments of 20.
These ranges were selected to correspond to the most frequently observed variations relative to the average price level of the S\&P 500 Index, thereby ensuring that the analysis captures empirically relevant dynamics rather than extreme or atypical cases (Figure \ref{fig: scatter}). For these analyses, we set the current price $P_t$  equal to 1333, corresponding to the average value of the S\&P 500 Index over the sample period considered in this study. In the Appendix \ref{appendix: current price}, we test the robustness of the results by varying the actual price. 

\begin{figure}[h]
\centering

    \centering
    \includegraphics[scale=0.5]{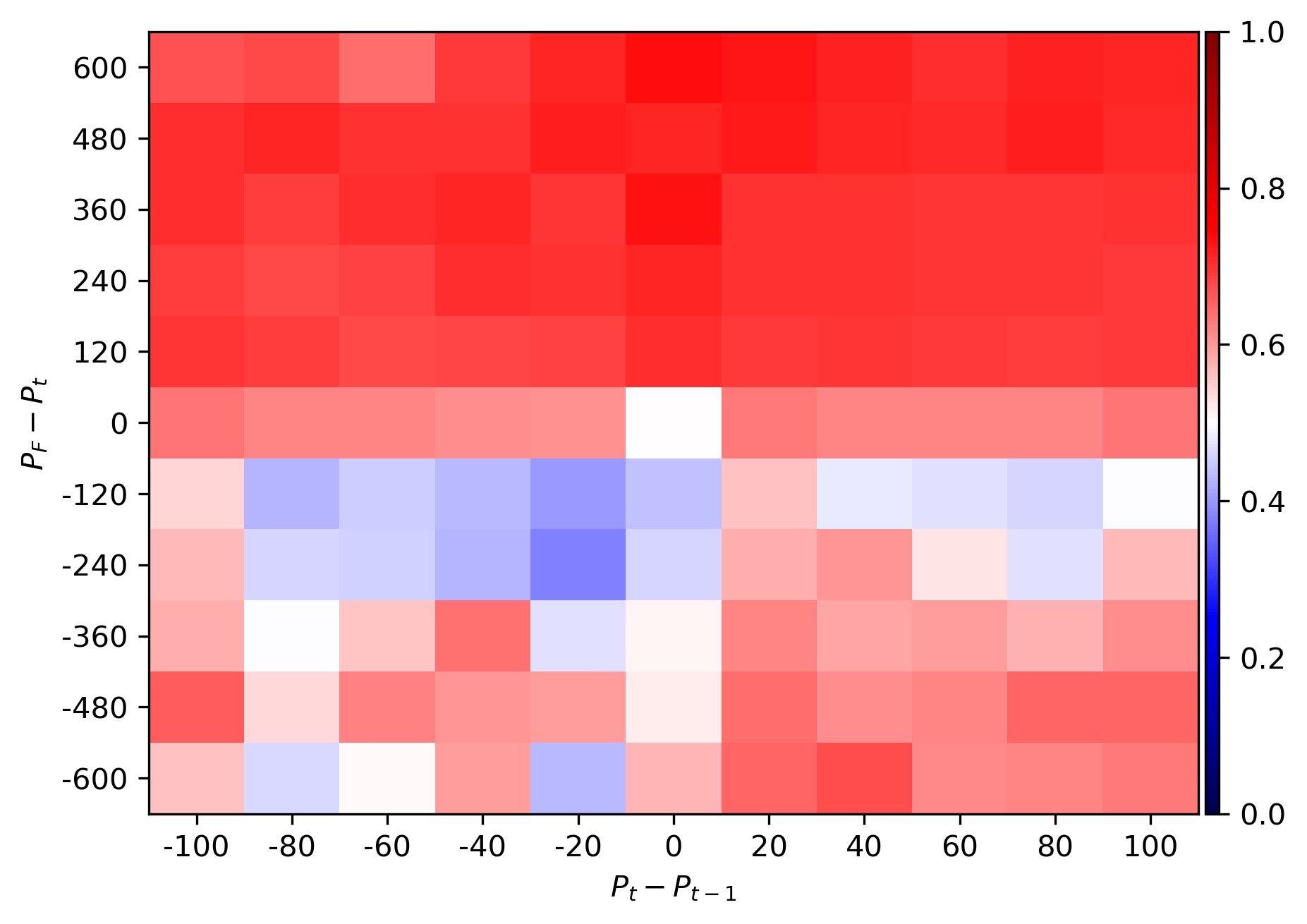}
    \caption{Probabilities of expectation adopted by the generative agent.}
    \label{fig:rules}

\label{fig:combined}
\end{figure}

For each combination of the current price, the previous price, and the fundamental value, we conduct 50 independent replications of Prompt 1 to simulate the market composition. The final results are obtained by averaging across the 50 replications. All simulations employ the \textit{gpt-4o-mini} model with a temperature parameter set to 1.
Figure \ref{fig:rules} shows the probability of assuming a fundamentalist strategy as a function of the deviation from the fundamental value $(P_t^f - P_t)$ on the vertical axis and the deviation from the lagged price $(P_t - P_{t-1})$ on the horizontal axis. Red squares indicate the prevalence of fundamentalists, conversely blue squares indicate the prevalence of trend followers. The results indicate that the generative agent exhibits distinct behavioral patterns.
When the market price is lower than the fundamental value ($P_t \leq P_t^f$), the fundamentalist strategy dominates the market, regardless of whether the market is in an upward trend (first quadrant) or a downward trend (second quadrant). This result  is consistently
replicated across all robustness check. In other words, regardless of whether the two strategies offer conflicting signals or move in the same direction, the generative agent adopts a fundamentalist expectation. In contrast, when the market price exceeds the fundamental price, blue colors appear and the probability of adopting a trend-following strategy increases, particularly when $P_t > P_t^f$ and $P_t < P_{t-1}$. Under this condition, agents extrapolate recent trends, expecting the price decline to continue. Interesting, when the market price is above the fundamental and the trend is increasing ($P_t > P_t^f$ and $P_t > P_{t-1}$), the greater the difference between the fundamental value and asset price, the higher the probability of choosing the fundamentalist strategy. This strategy mirrors the one introduced by \citep{westerhoff2009exchange}, which assumes an increasing adoption of the fundamentalist strategy when prices deviate excessively from the fundamental value, indicating an unstable trajectory. In fact, as noted by \citep{westerhoff2009exchange}, the greater the discrepancy between the fundamental price and the market price, the stronger the belief in mean reversion.

To summarize the results obtained, simulations based on artificial market data confirm the presence of heterogeneity in expectations. At the same time, the prevalence of red areas indicates that fundamentalist expectations prevail across the different cases, particularly when the fundamental price exceeds the actual price. Conversely, blue regions emerge when the actual price lies below the fundamental or when recent price changes are negative, suggesting that trend follower behavior becomes more likely under conditions of positive mispricing and downward momentum. 

\begin{figure}[ht]
    \centering
    % Riga 1
    \includegraphics[width=0.49\textwidth]{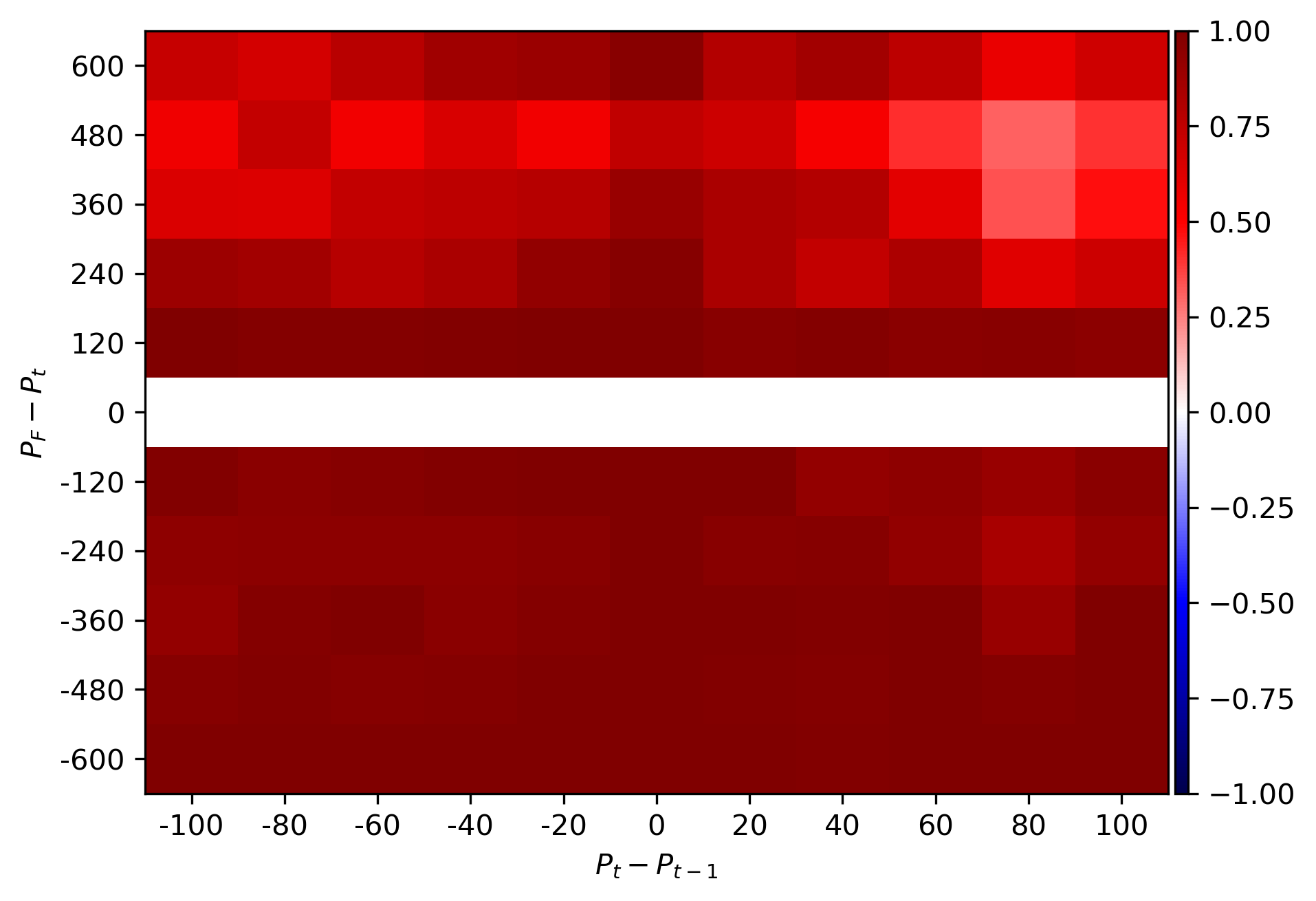}\hfill
    \includegraphics[width=0.49\textwidth]{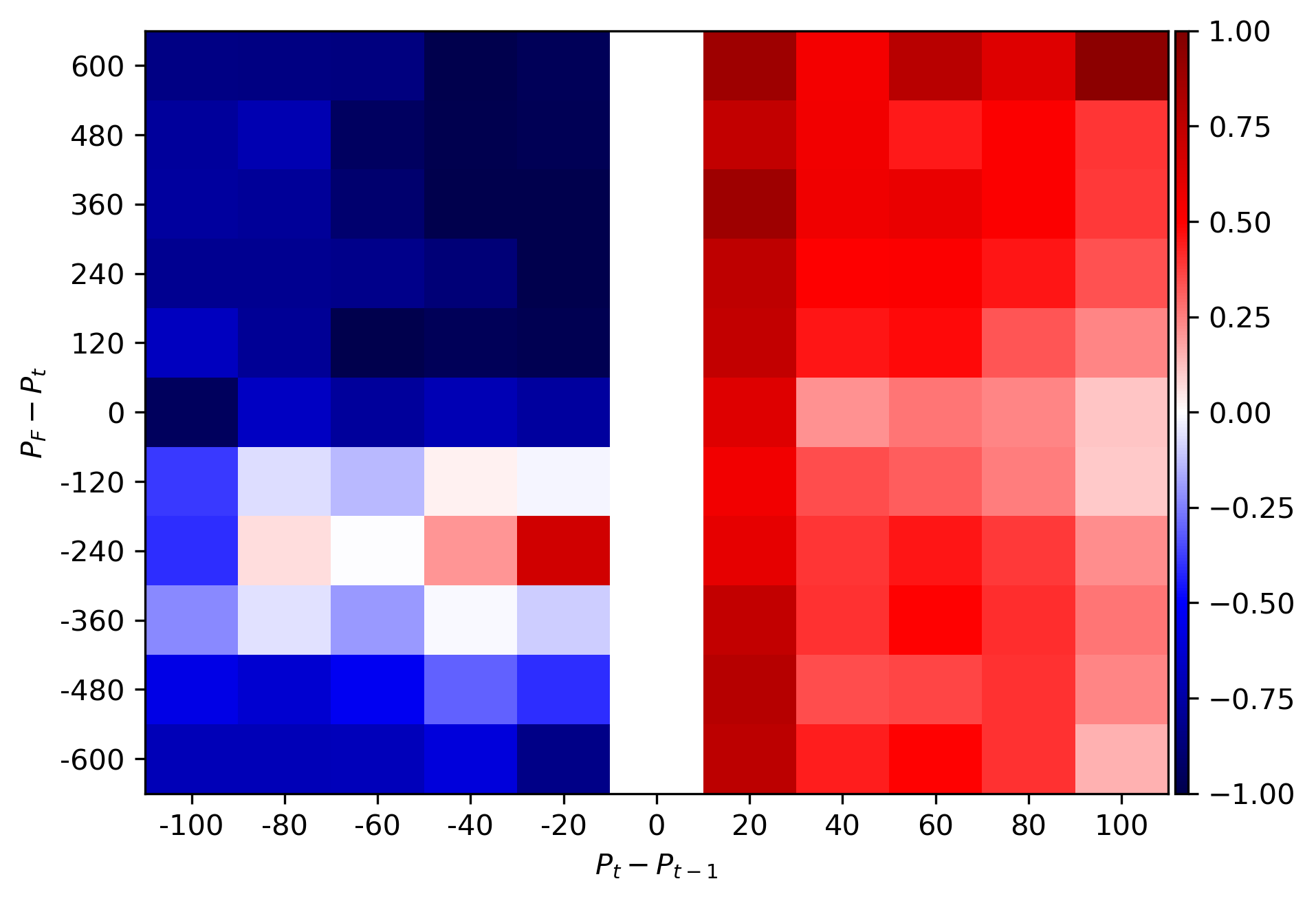}\\[1ex]
    % Riga 2
    
    \caption{Average values of the reaction parameters for fundamentalists (on the let) and for trend followers (on the right).}
    \label{fig:reaction_param}
\end{figure}

At this point of the analysis, we can extract the reaction parameters implied by the expectations of the generative agent. To this end, we simulate Prompt 2 and use the combination of price and fundamental value, as in the previous exercise. Based on 50 Monte Carlo simulations, we use Eq. \ref{eq: alpha} and Eq. \ref{eq: beta} to extract the average values of the reaction parameters $\alpha$ (for the fundamentalist expectation) and $\beta$ (for the trend follower expectation). 
In Figure \ref{fig:reaction_param},
we present the results obtained when a fundamentalist expectation is used (on the left) and when trend follower behavior is chosen (on the right).

As shown in the figure on the left, for the fundamentalists, the value of the reaction parameter is consistently positive and lies between zero and one, as suggested by theory. This implies that the generative agent always pushes in the direction of the fundamental value, never exhibiting contrarian behavior with respect to it. In contrast, for generative trend follower, both positive and negative values of the reaction parameter are observed. When the price lies below the fundamental, two qualitatively different states can be
observed. In periods of upward trends, the reaction parameter is strongly positive, implying
that the trend follower expects the appreciation to continue toward the fundamental. Conversely, during downward trends, the reaction parameter becomes negative, indicating that the
trend follower behaves as a contrarian, anticipating a trend reversal, and once again expecting
an appreciation toward the fundamental. Thus, when the price is below its fundamental value,
both fundamental and trend followers contribute to pushing the expected price upward. This
pattern is robust in all the variations tested.  
              
Finally, we combine the results obtained from Prompt 1, which identifies the composition of the market, with those of Prompt 2, which generates the agents' expectations, to calculate the expected price for the following period, as previously illustrated in Equation \ref{eq: price_exp}. We derive the expected price variation ($E\left[\Delta(P_{t+1}) \right]$) by subtracting the current price from the expected price:

\begin{equation}\label{eq: expected price variation}
    E\left[\Delta(P_{t+1}) \right] = E[P_{t+1}]-P_t= \delta_t \alpha_t (P^f_t-P_t)+(1-\delta_t)\beta_t(P_t-P_{t-1})
\end{equation}

\begin{figure}[hb]
\centering

    \centering
    \includegraphics[scale=0.5]{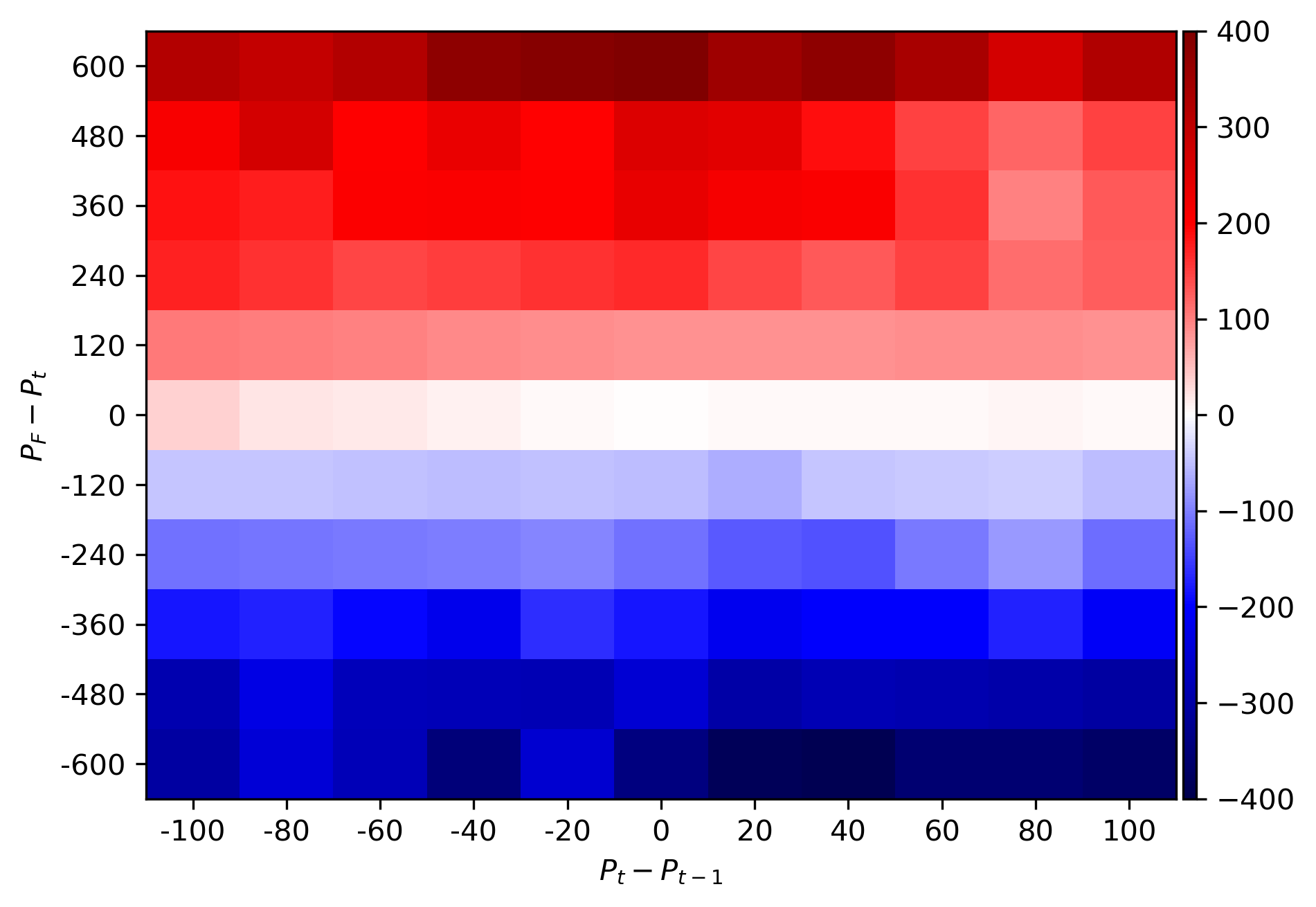}
    \caption{Expected price variation.}
    \label{fig: expected price variation}

\label{fig:combined}
\end{figure}

The resulting values are presented in Figure \ref{fig: expected price variation}.

As shown, the expected price variation is positive when the fundamental value exceeds the current price and negative when the fundamental is below the current price. Moreover, the absolute magnitude of the expected variations increases with the distance between the current price and the fundamental value, while no significant change is observed with respect to variations in the difference between the current and previous prices. This outcome is fully consistent with efficient market theory, according to which prices are expected to converge toward their fundamental values.
This result can be explained by two main factors. First, the deviations between the fundamental value and the current price are substantially larger than those between the current and previous prices, thus assigning greater weight to the fundamental component in Equation \ref{eq: expected price variation}. Second, the reaction parameter ($\alpha$) of the fundamentalist agents, which takes values close to one, further amplifies the influence of the fundamentalist strategy in determining the expected future price.

\section{Conclusions}

Large language models enable social scientists to move beyond conventional methodologies and pursue innovative research approaches. This paper explored the intersection between large language models and economics, with the aim of assessing whether LLMs can replicate the heuristic expectation-formation mechanisms described in the heterogeneous agent modeling literature, in which market dynamics arise from the coexistence of agents holding different beliefs. 

In our analysis, we employed the ChatGPT \textit{gpt-4o-mini} model, accessed via a Python script, and interacted with the LLM using prompting techniques. The study focused on two main aspects: first, the composition of the market, and second, the expectations of the generative agents according to their respective belief systems. Both prompts are tested on real time series data and on artificial market data. 

In the time-series analysis, both behavioral strategies formalized in HAMs emerge as relevant, with a higher probability of trend-following behavior preceding the dot-com and global financial crises. This is fully consistent with the predictions of heterogeneous agent models. At the same time, mixed results are obtained during the bust phases. In fact, following the 2007–2008 financial crisis, the adjustment phase is driven by fundamentalist behavior, whereas a negative trend following pattern dominates in the aftermath of the dot-com crisis. Furthermore, when explicitly modeling agents’ expectations, the estimated reaction parameters lie between 0 and 1 for fundamentalists and take both positive and negative values for trend followers. 

In the artificial market simulations, both strategies are again identified, although generative fundamentalist behavior prevails. Moreover, the analysis of market composition reveals a marked asymmetry between cases in which the current price is above or below the fundamental value. For prices above the fundamental, the generative agent is likely to switch between the two alternatives. At the same time, for the generative agent, there appears to be no justification for the price to remain below its fundamental value. In fact, when the price lies below its fundamental value, the generative agent systematically selects the fundamentalist strategy.  This interpretation is further supported by the analysis of the reaction parameters derived from the expectations of the generative trend-following agents. 

In summary, the LLM successfully reproduces a market composition broadly consistent with the predictions of the heterogeneous agent modeling literature with filtered reaction parameters in the range predicted by the HAM literature. At the same time, artificial market simulations provide additional insight into the behavioral rules of the generative agent filtered through the information contained in the prompts. First, some of the heuristics underlying these results lack theoretical support, particularly with respect to the qualitatively asymmetric behavior that depends on the relative position of the fundamental value and the current price.  Second, the $\lambda$ parameter being close to unity and the predominant red-shaded area in the probability of selection of the strategy both point to a greater relevance of fundamentalist behavior. In support of this, by combining the composition of the market with the expectations of the agents, we derived the expected prices implied by a market composed entirely of generative agents. The resulting expected price converges toward the fundamental value, regardless of whether the market exhibits an upward or downward trend.       

In the late 1970s, Simon argued that economics should move beyond the narrow confines of substantive rationality and embrace a theory of procedural rationality, capable of capturing actual decision making under bounded information and computational resources. Crucially, he anticipated the relevance of cross-disciplinary contributions, including artificial intelligence, to advance this research agenda \citep{simon1978rationality}. In line with this vision, recent advances in artificial intelligence, particularly the emergence of large language models, have further expanded the methodological frontier of the social sciences. LLMs are increasingly employed as flexible research tools capable of supporting diverse analytical applications, ranging from qualitative text analysis and survey design to discourse modeling and theory testing. Their adaptability allows researchers to tailor them to specific empirical contexts, while their scalability facilitates large-scale analyses that would be prohibitively costly or time-consuming using traditional approaches \citep{bail2024can, bilancini2024ai,zhang2024chatbot,argyle2025arti}. As such, LLMs not only exemplify the integration of computational methods into social inquiry but also embody Simon’s call for a more procedural and interdisciplinary understanding of human decision-making.

\bibliographystyle{chicago}

\newpage
\appendix

\section{Robustness input}
In this appendix, we perform robustness checks by varying the inputs provided to the LLM. In subsection \ref{appendix: prompt}, we vary the prompt, while in subsection \ref{appendix: current price} we modify the current price, which also affects the fundamental value and the price of the previous period. For price modification, the analysis is repeated both for strategy identification (first script) and for expectation formation (second script). Finally, we test the robustness of the results by increasing the number of repetitions in subsection \ref{appendix: dimension}.

\subsection{Prompt}\label{appendix: prompt}

Through interactions with ChatGPT, we generate five variations of the original prompt. These variations preserve its overall meaning while altering the sentence structure and replacing some terms with synonyms. For each version, we run 10 repetitions with the current price set to 1333, the temperature set to 1, and using the model gpt-4o-mini. 

\begin{figure}[H]
\centering
    \includegraphics[scale=0.49]{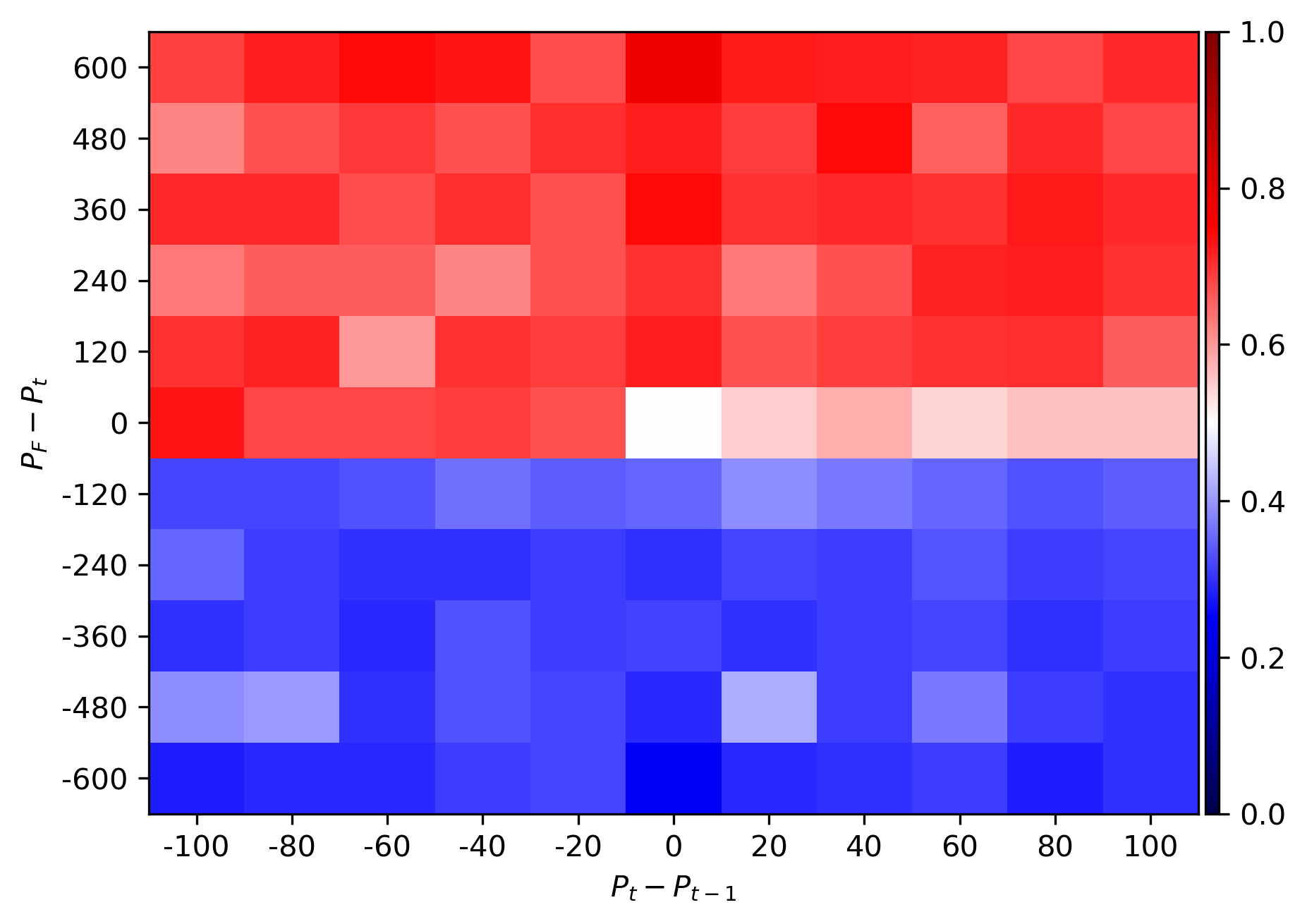}
    \includegraphics[scale=0.49]{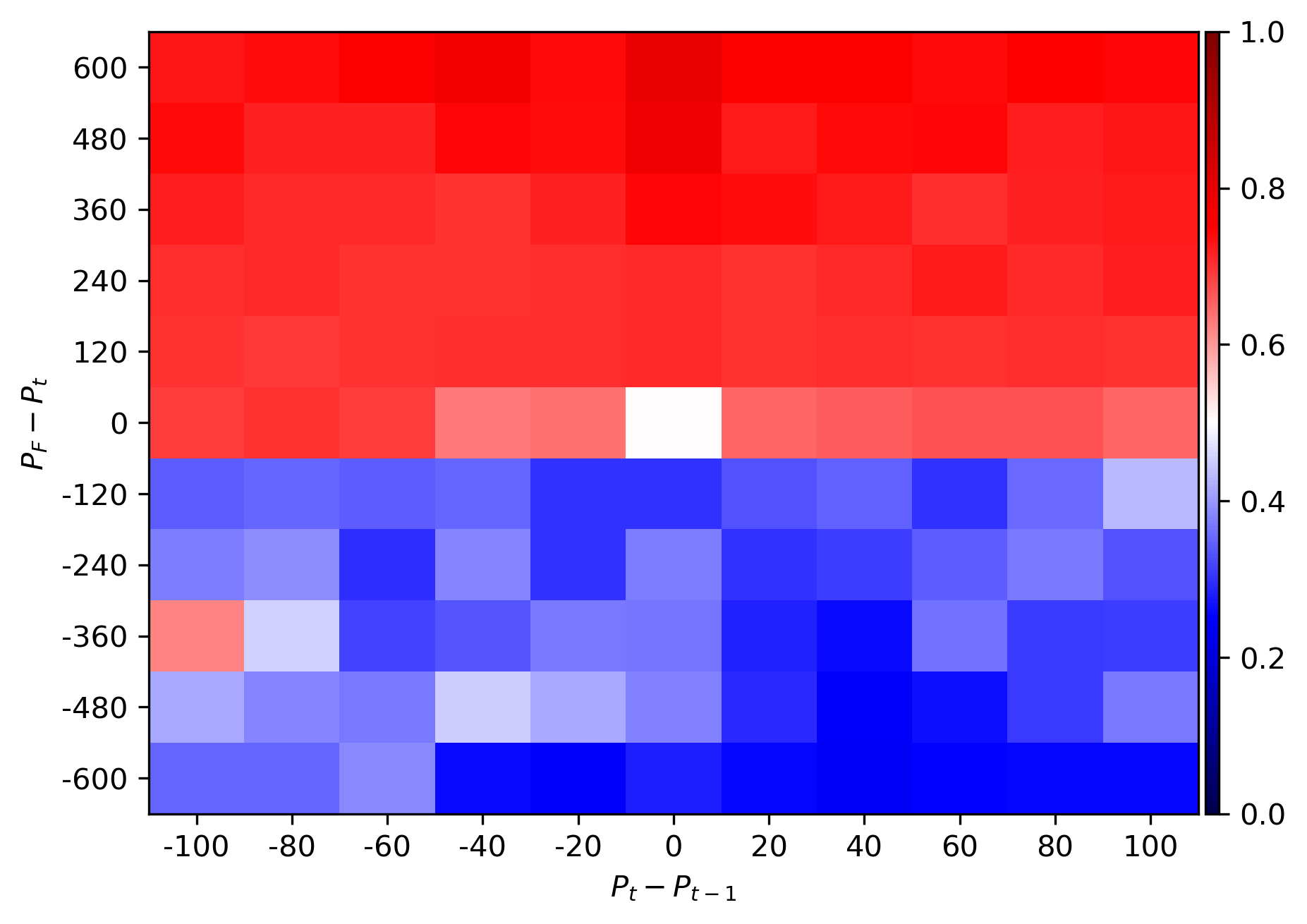}
    \includegraphics[scale=0.49]{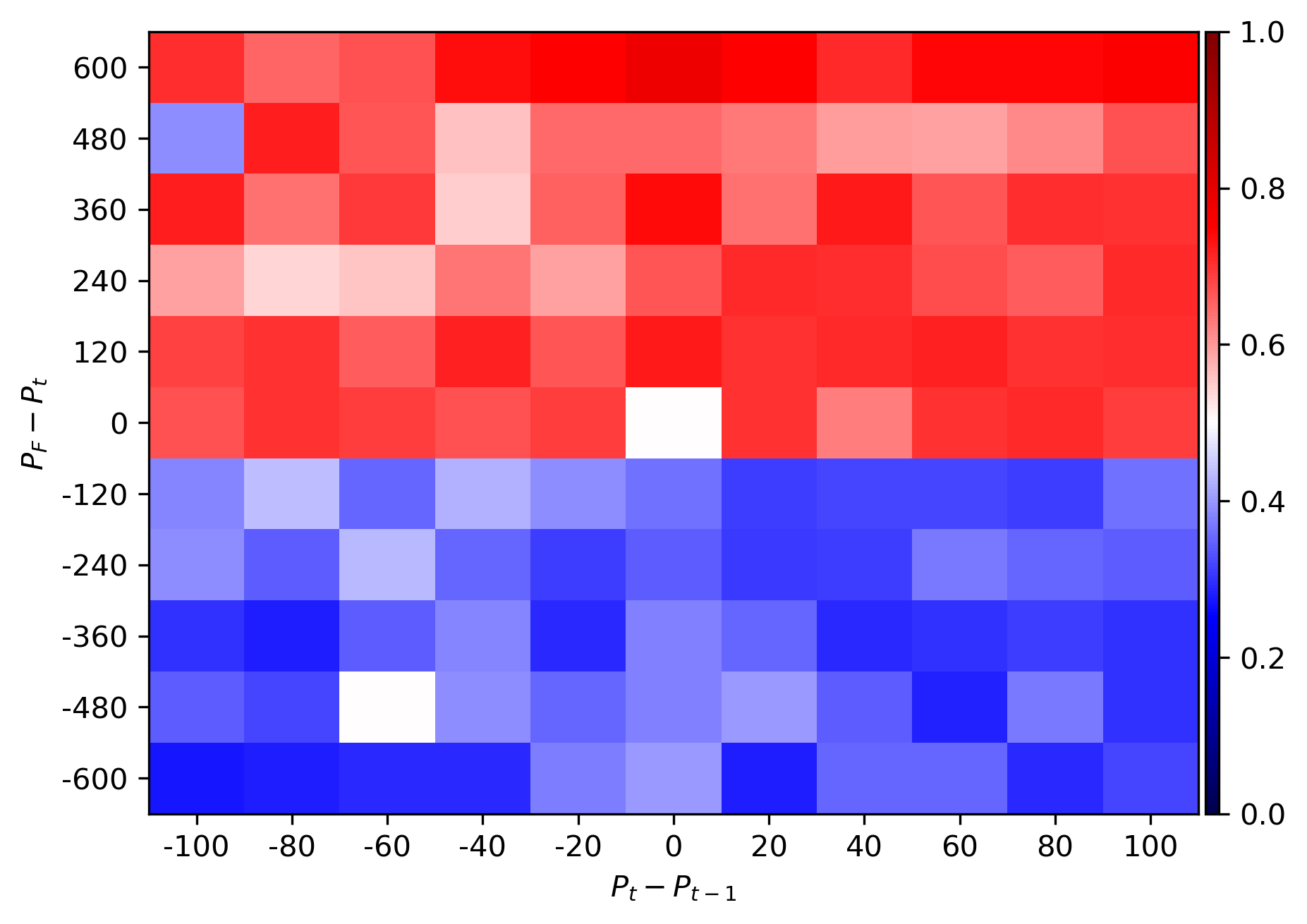}
    \includegraphics[scale=0.49]{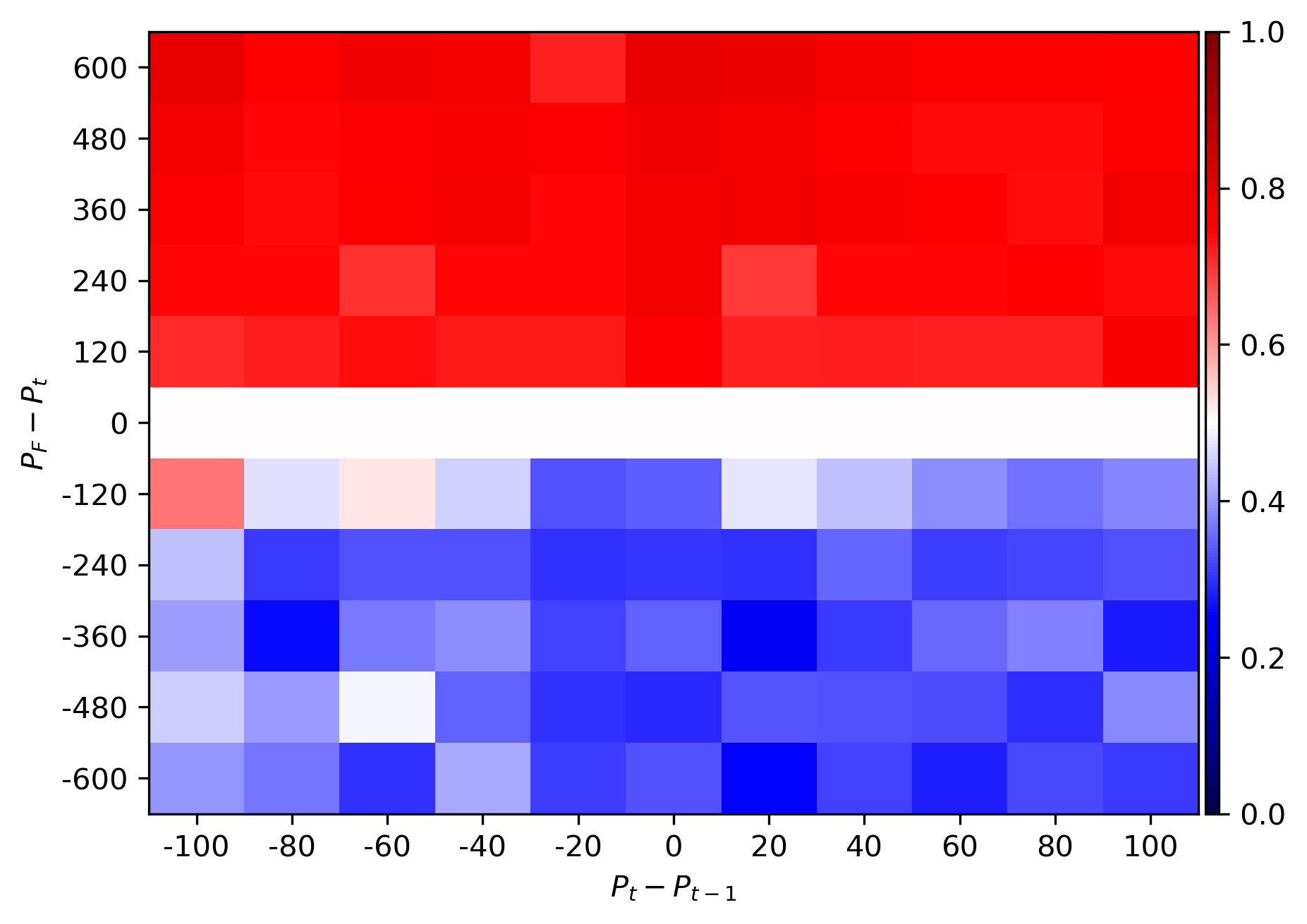}
    \includegraphics[scale=0.49]{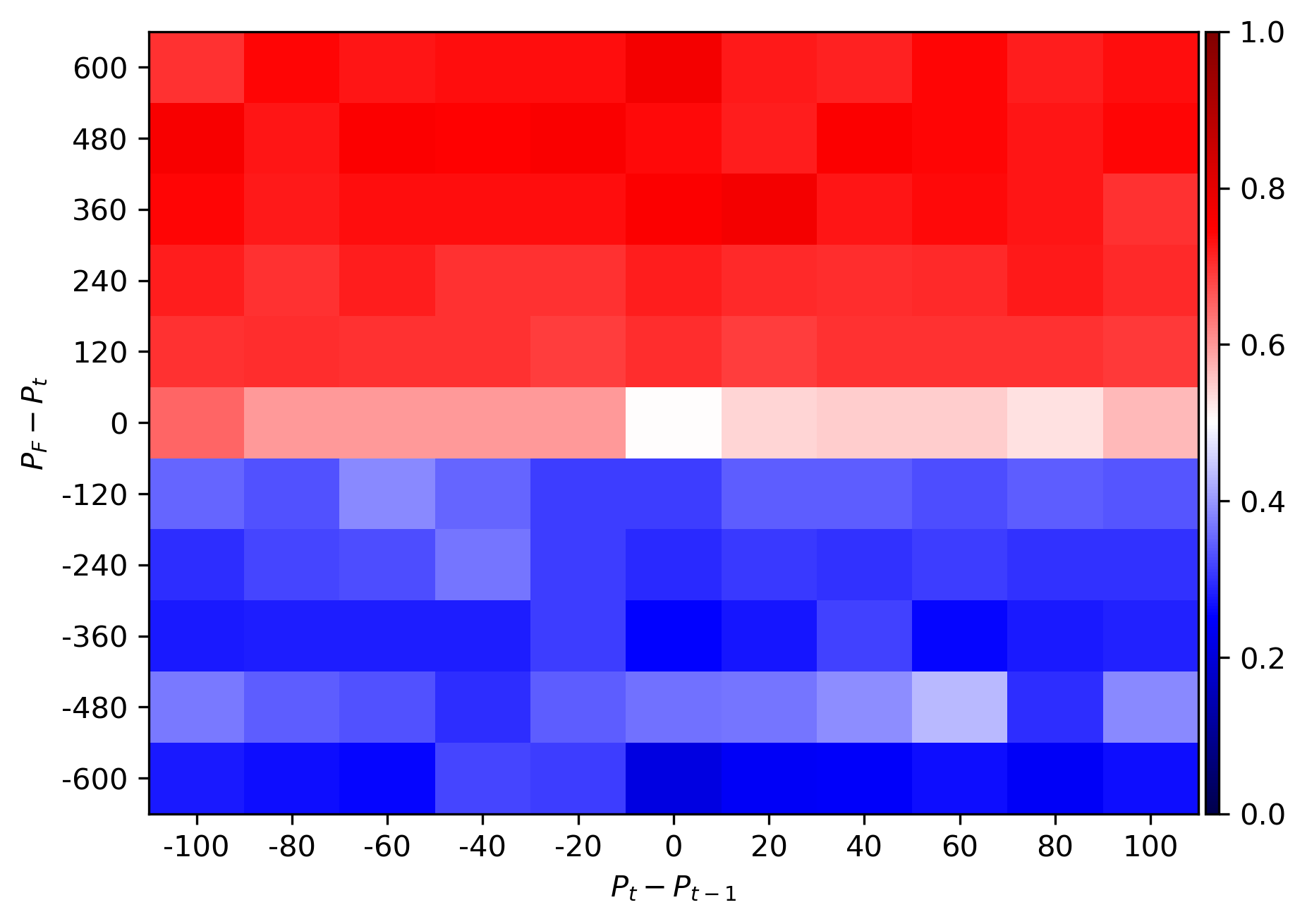}
    \caption{Weight of the fundamentalist strategy. Results are averaged over 10 repetitions with \textit{gpt-4o-mini} at temperature 1 and $P_t=1333$. Top left: \textbf{Version 1} of the prompt; Top right: \textbf{Version 2} of the prompt; center left: \textbf{Version 3} of the prompt; center right: \textbf{Version 4} of the prompt; bottom: \textbf{Version 5} of the prompt.}
    \label{fig:prompt1}
\end{figure}

Anticipating the results, prompt variations confirm the asymmetry between periods when the price is above versus below the fundamental. As before, fundamentalists dominate when the price is below the fundamental, while trend followers increase when the price is above it. The prompts are reported below and the results are in Figure \ref{fig:prompt1}. \\

\textbf{Version 1:} \textit{You are acting as a market trader who must decide between two strategies based on expectation formation: a fundamentalist strategy and a trend-following strategy. You are given the current fundamental value, the current asset price, and the asset price from the previous period. Your task is to output only two weights, representing the probabilities assigned to each strategy. The weights must sum to 1. Respond only with the two weights separated by a comma, with no additional text.} \\

\textbf{Version 2:} \textit{Imagine you are a financial trader selecting between two possible trading strategies: fundamentalist or trend-following. Your choice should be guided by expectation formation and the data provided (fundamental value now, current price, and past period price). For each strategy, assign a probability weight. The two weights must add up to 1. Your response must include only the two weights separated by a comma, nothing else.} \\

\textbf{Version 3:} \textit{As a trader, you are tasked with choosing between two market approaches: the fundamentalist strategy and the trend-following strategy. You base your decision on the given information: the current fundamental value, the current asset price, and the asset price from the prior period. Provide only two numbers, each representing the probability of selecting one of the strategies. These numbers must sum to 1. Your reply should contain only the two weights separated by a comma.} \\

\textbf{Version 4:} \textit{You are a trader deciding between two alternative strategies informed by expectation formation: (1) a fundamentalist approach and (2) a trend-following approach. The available data are the current fundamental, the present asset price, and the asset price from the last period. Output exactly two weights—probabilities for each strategy—that sum to 1. Provide only the two weights separated by a comma, with no extra commentary.} \\

\textbf{Version 5:} \textit{Take the role of a financial trader who must allocate probabilities between two strategies: a fundamentalist strategy and a trend-following strategy. Use the given information (current fundamental, current price, and last period’s price) to form expectations. Output only two values representing the weights for the two strategies. They must sum to 1. Your answer should consist only of the two weights separated by a comma.} \\

As an additional robustness check, we modify the prompt to simultaneously obtain both the weights associated with the two strategies and the expectations linked to them. Instead of using the two separate prompts presented in the main analysis, we employ the following prompt: \\

\textbf{Prompt:} \textit{You are a financial trader and you need to choose between two alternative market strategies according to your analysis based on expectation formation: the first is a fundamentalist approach, and the second is a trend-following approach. You have access to the current value of the fundamental, the current value of the asset price, and the value of the asset price at the previous period. You have to provide a weight that represents the probability with which you would choose the strategy. The sum of the weights must be 1. Do not provide any other comment, just the two weights divided by a comma and included in double quotes each one. Moreover, you need to provide the value of price expectations for the next period using the two alternative market strategies. Do not provide any other comment or word, just the two numbers divided by a comma and included in double quotes each one. You must write the entire answer on a single line.} \\

\begin{figure}[h]
\centering
    \includegraphics[scale=0.48]{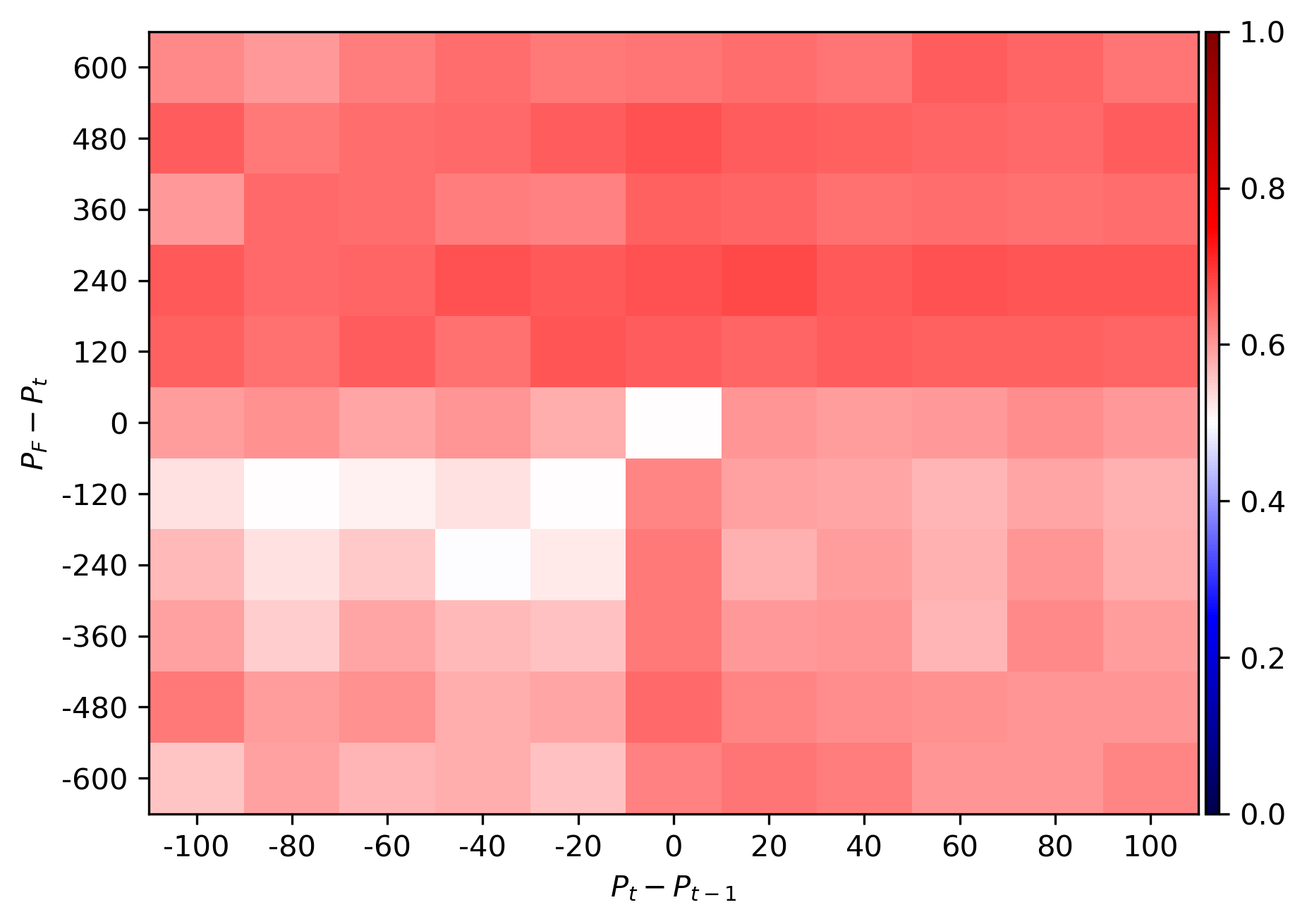}
    \includegraphics[scale=0.48]{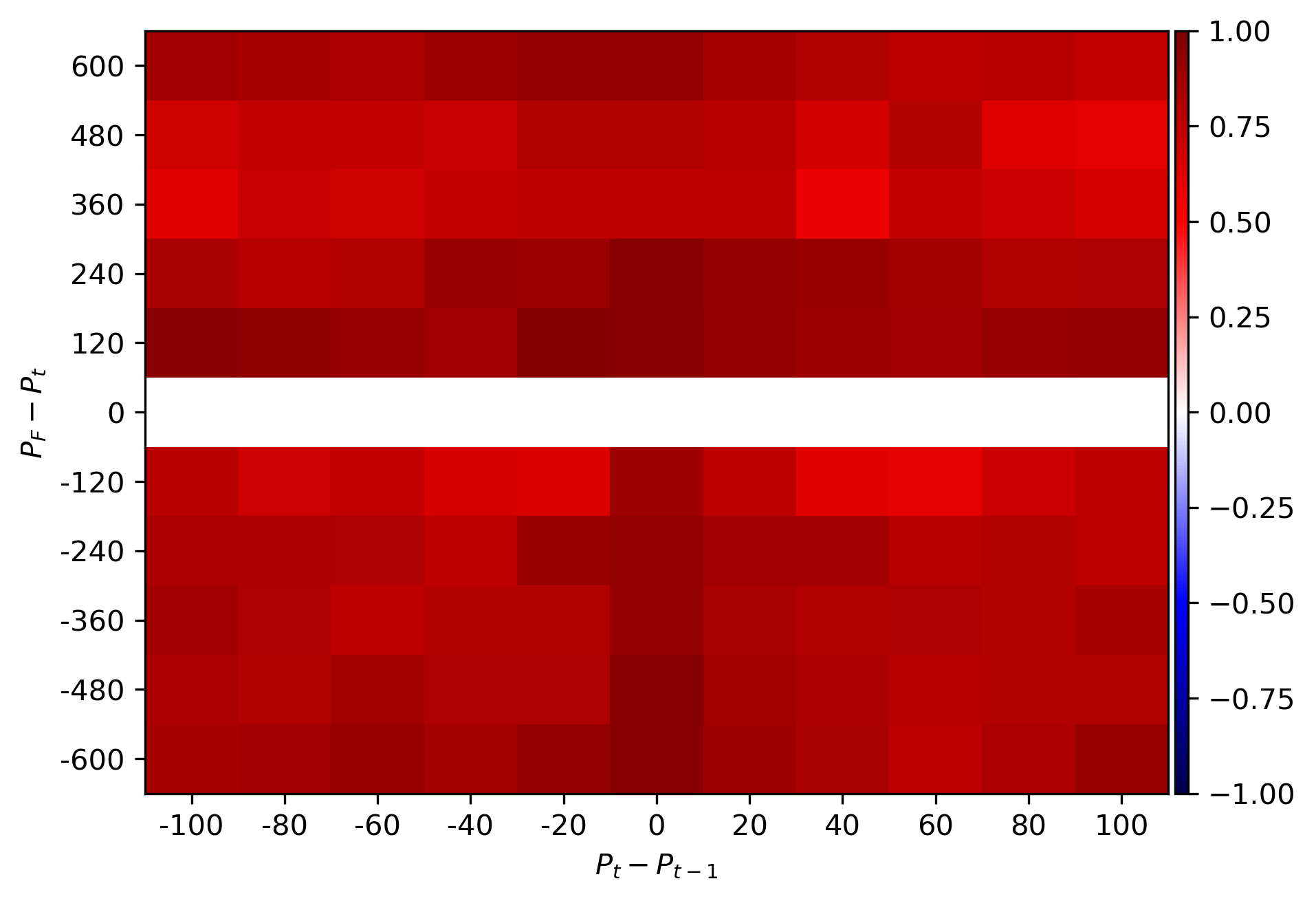}
    
    \includegraphics[scale=0.48]{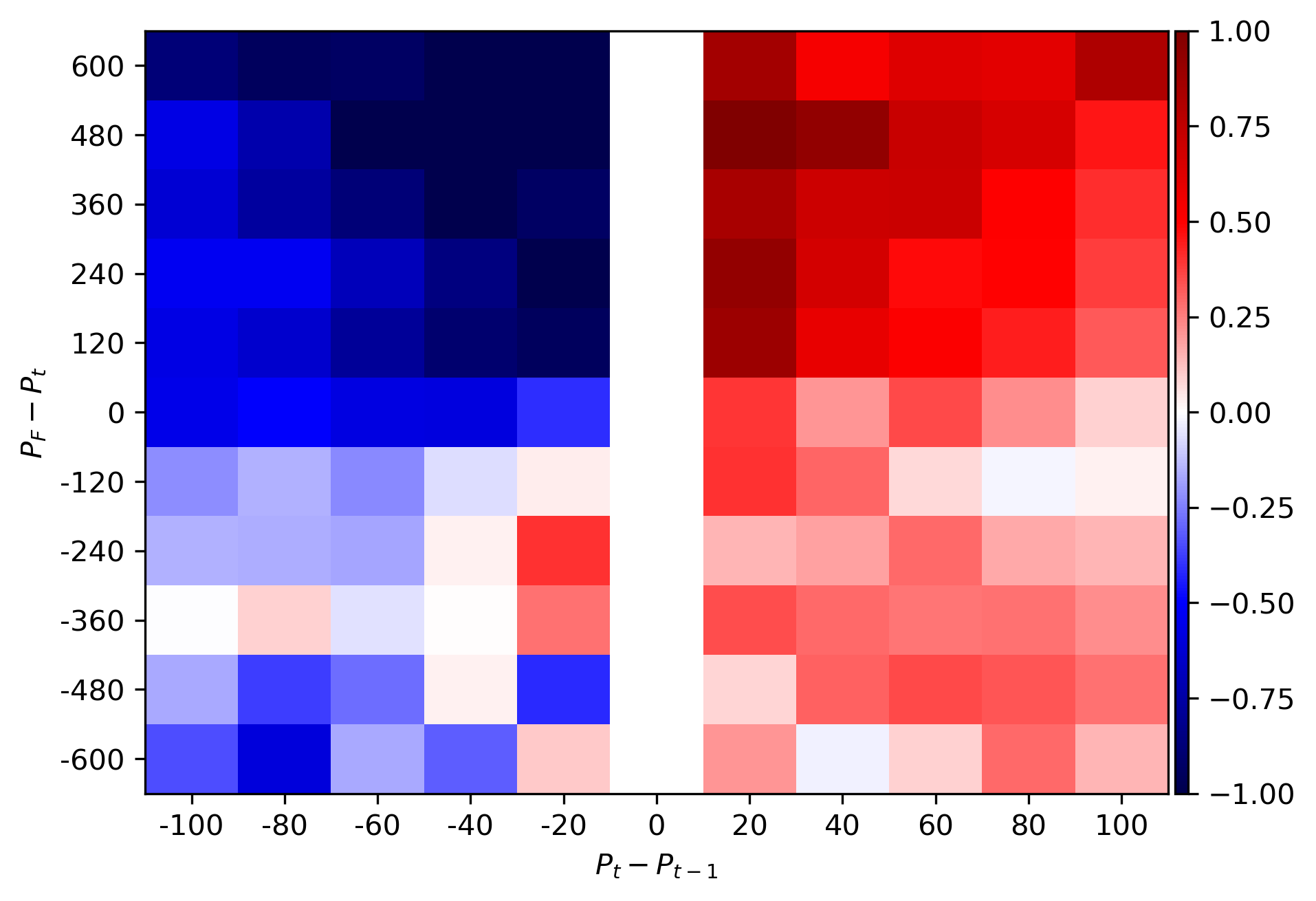}    \includegraphics[scale=0.48]{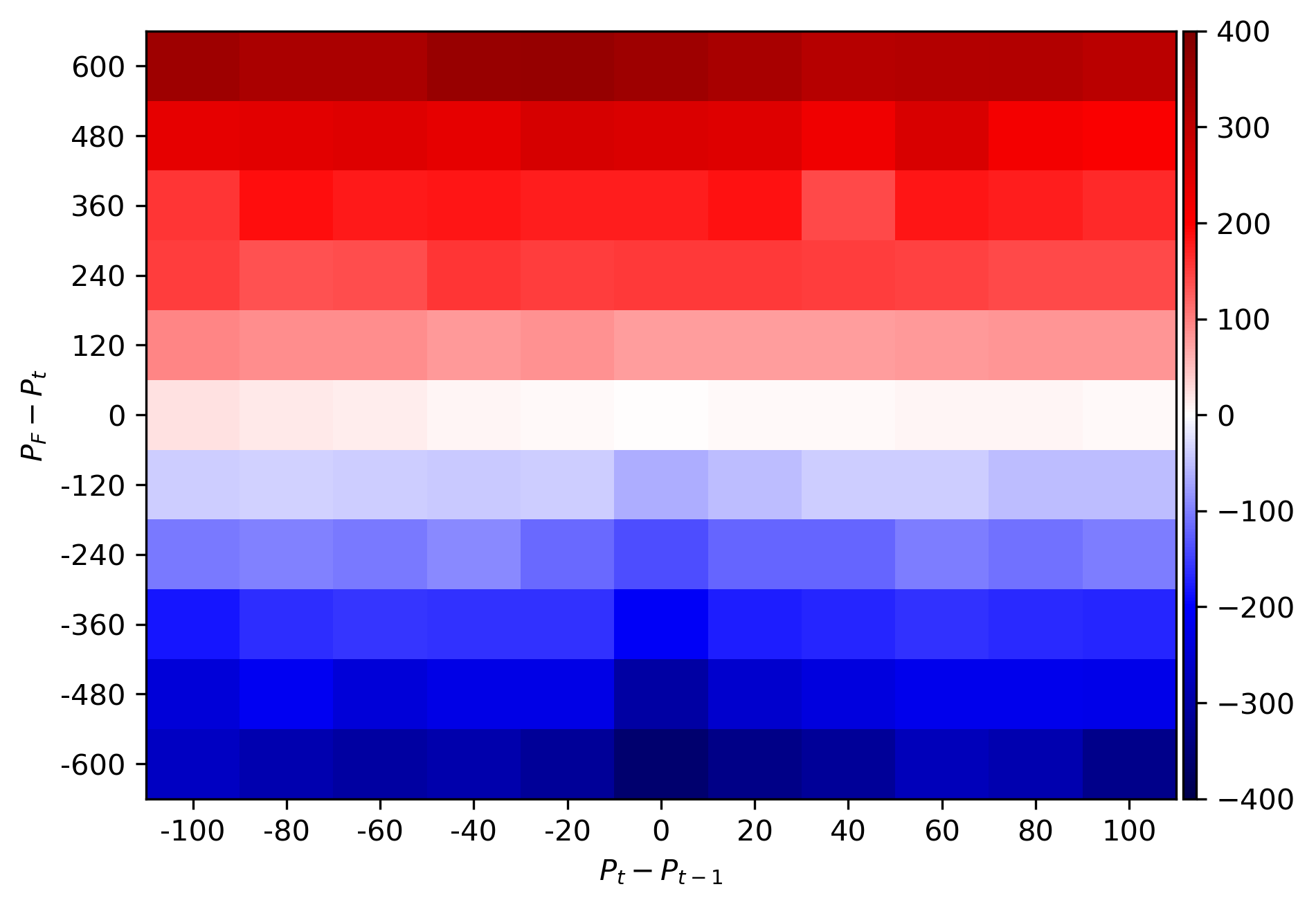}
    
    \caption{Top left: weight of the fundamentalist strategy; Top right: $\alpha$; Bottom left: $\beta$ parameter; Bottom right: Expected price variation. Results are averaged over 50 repetitions with \textit{gpt-4o-mini} at temperature 1 and $P_t=1333$, using the unified prompt.}
    \label{fig:prompt unified}
\end{figure}

As shown in Figure \ref{fig:prompt unified}, the graphs for the values of $\alpha$ and $\beta$ closely resemble those obtained in the main analysis. However, the plot for the relative weight of the fundamentalists shows some differences; nevertheless, the main result is confirmed: there is an asymmetry between periods when the price is above or below the fundamental value. Specifically, the predominance of fundamentalists when the price is below the fundamental decreases once the price moves above it. Finally, we present the results for the expected price variation, confirming the robustness of the findings when using an integrated prompt that jointly generates strategies and expectations.

\subsection{Current price}\label{appendix: current price}

\begin{figure}[ht]
    \centering
    % Riga 1
    \includegraphics[width=0.49\textwidth]{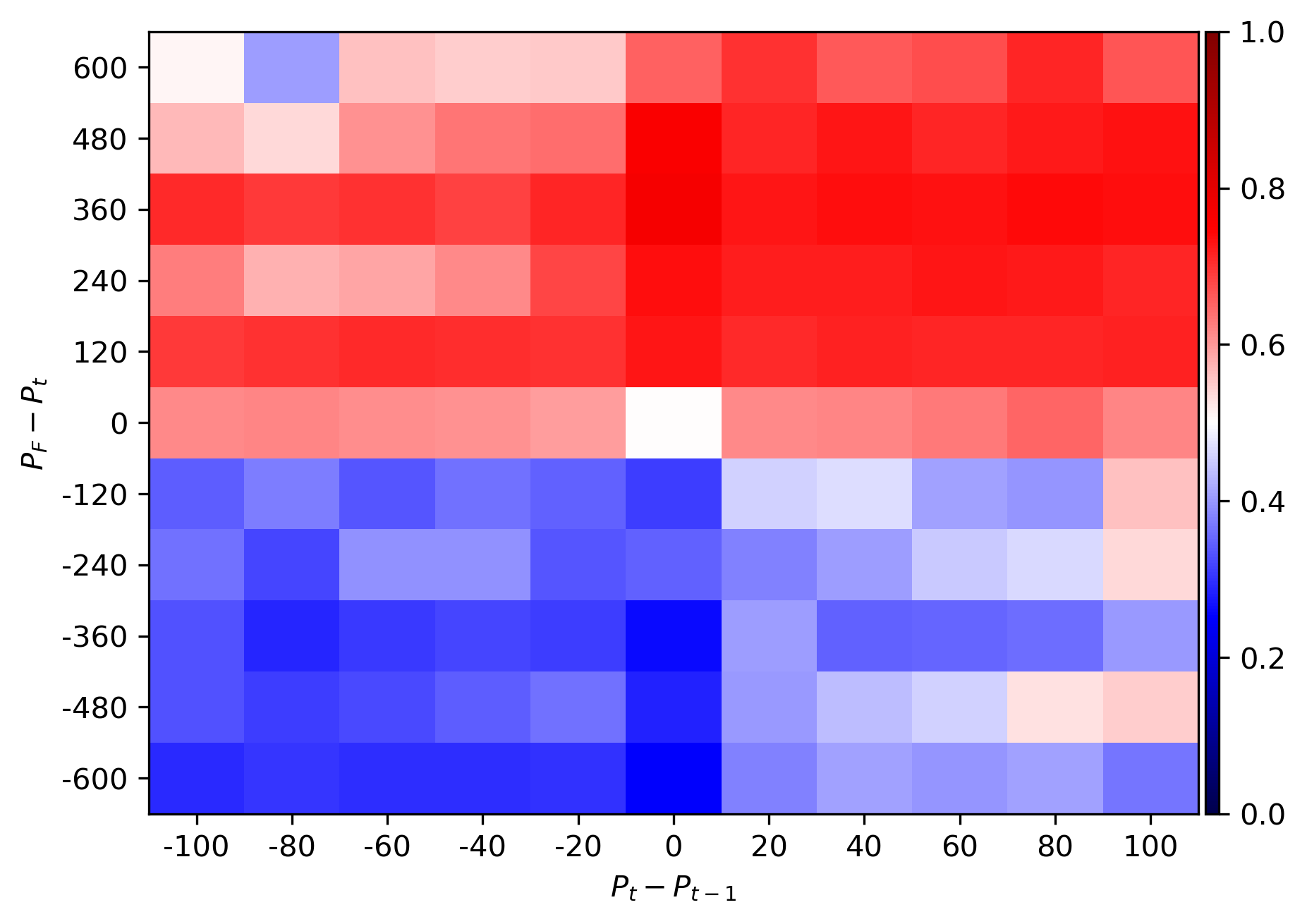}\hfill
    \includegraphics[width=0.49\textwidth]{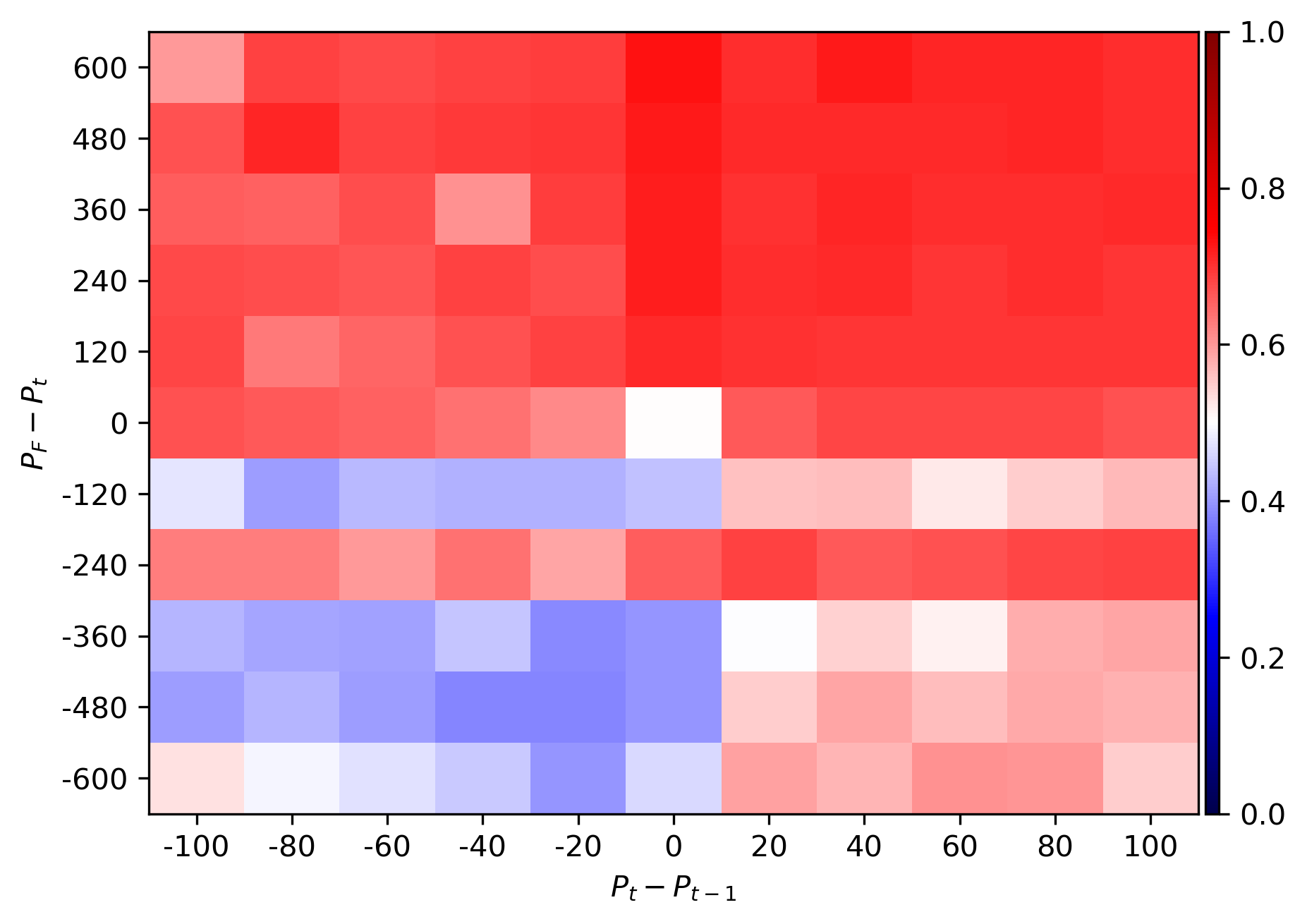}\\[1ex]
    % Riga 2
    \includegraphics[width=0.49\textwidth]{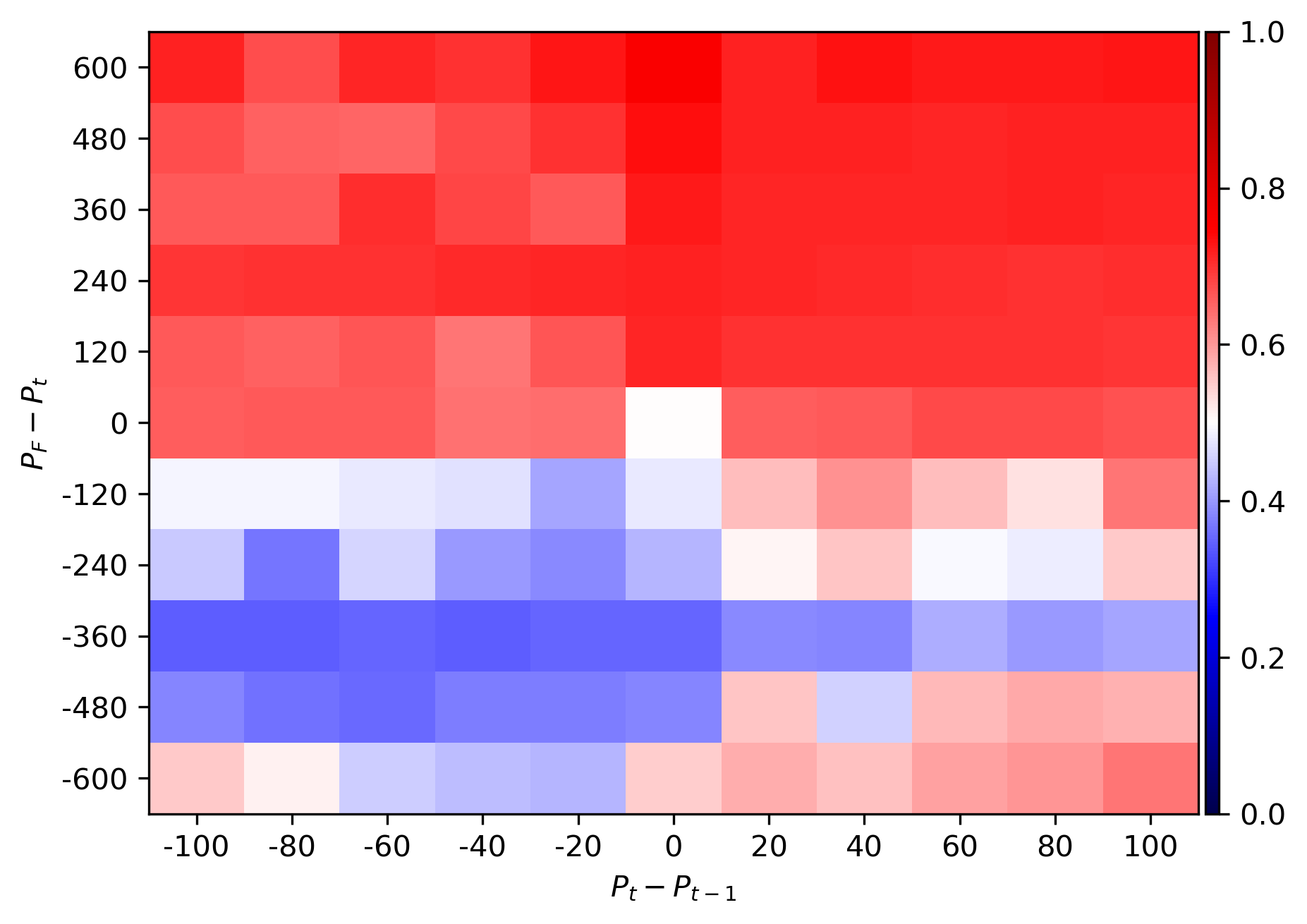}\hfill
    \includegraphics[width=0.49\textwidth]{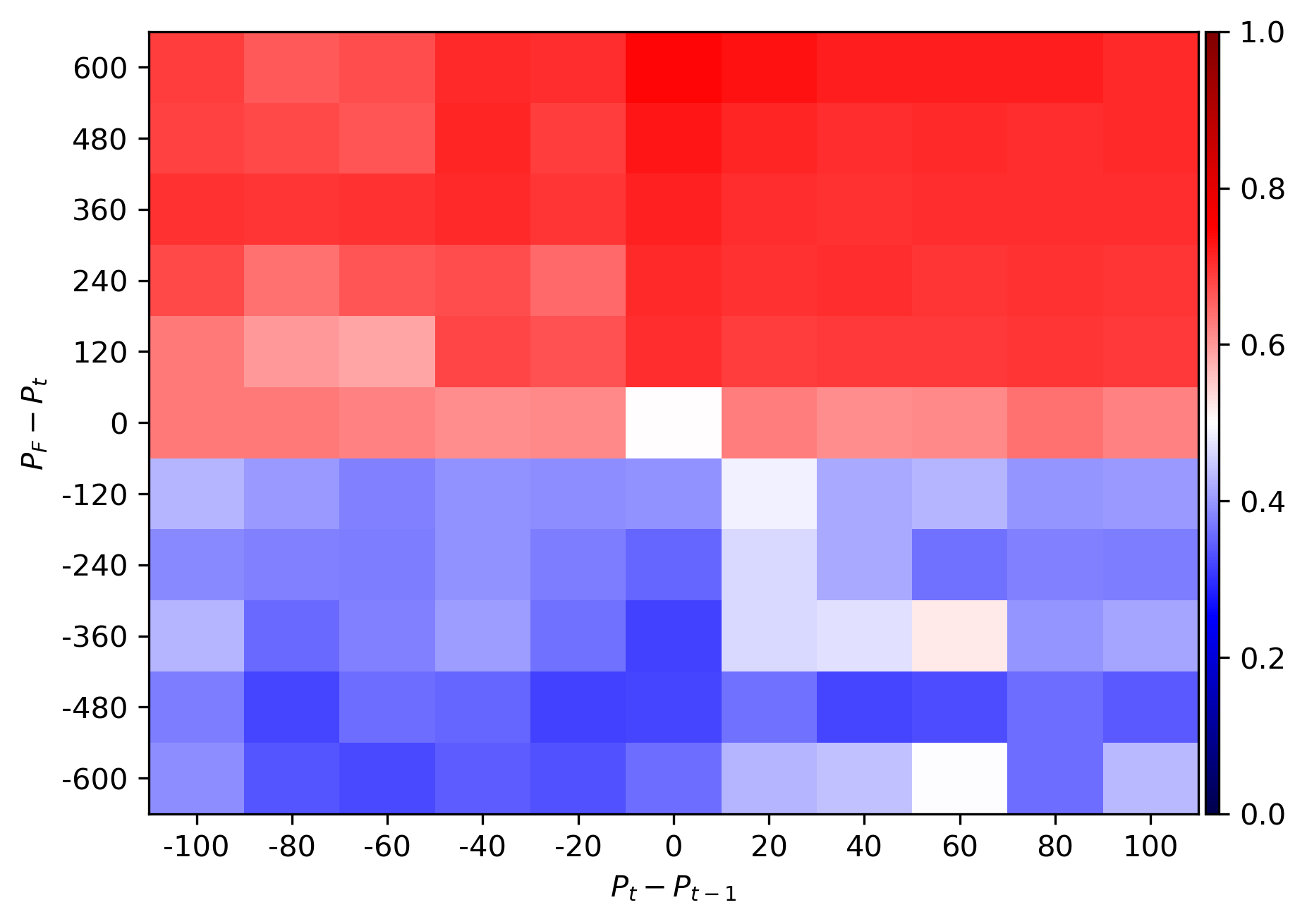}
    \caption{Weight of the fundamentalist strategy. Top-left: simulations for $P_t=833$; top-right: $P_t=1083$; bottom-left: $P_t=1583$; bottom-right: $P_t=1833$. Each subfigure reports averages over 50 repetitions with \textit{gpt-4o-mini} at temperature 1.}
    \label{fig: stretgie pt appendix}
\end{figure}

In this subsection, we vary the current price. To ensure comparability, we keep $P_F-P_t$ and $P_t-P_{t-1}$ constant; thus, changing $P_t$ also implies changes in $P_F$ and $P_{t-1}$. The new current price values are: 833, 1083, 1583, and 1833. All other parameters are kept constant, and we perform 50 repetitions for each set of values using \textit{gpt-4o-mini} with temperature 1. The results are replicated both for identifying the strategies used by the LLM and for determining expectations about the future price under the two strategies.

The simulation results for the weights assigned to the two strategies are reported in Figure \ref{fig: stretgie pt appendix}. Once again, we observe an asymmetry between cases where the current price is above or below the fundamental value.

\begin{figure}[H]
    \centering
    % Riga 1
    \includegraphics[width=0.48\textwidth]{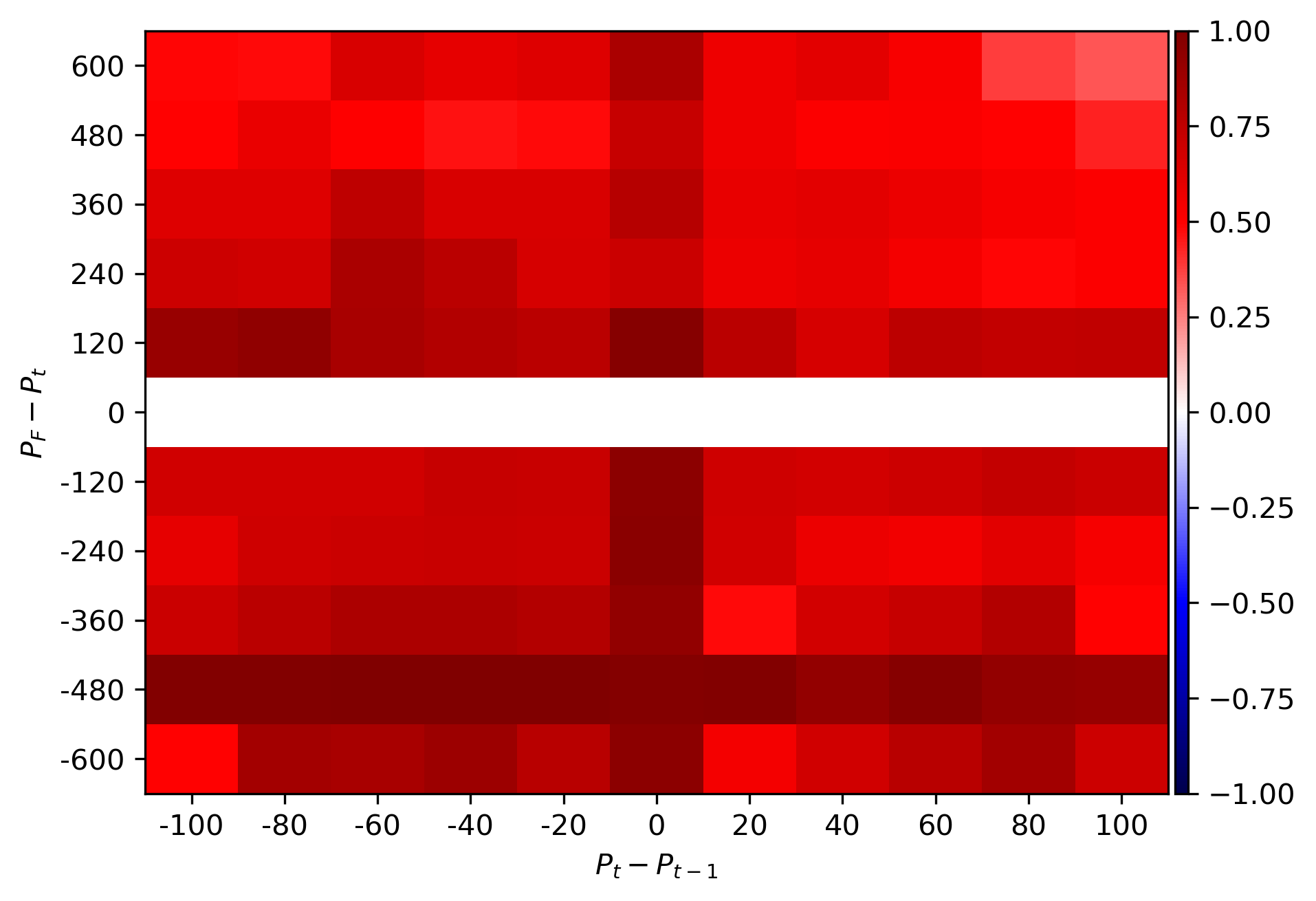}\hfill
    \includegraphics[width=0.48\textwidth]{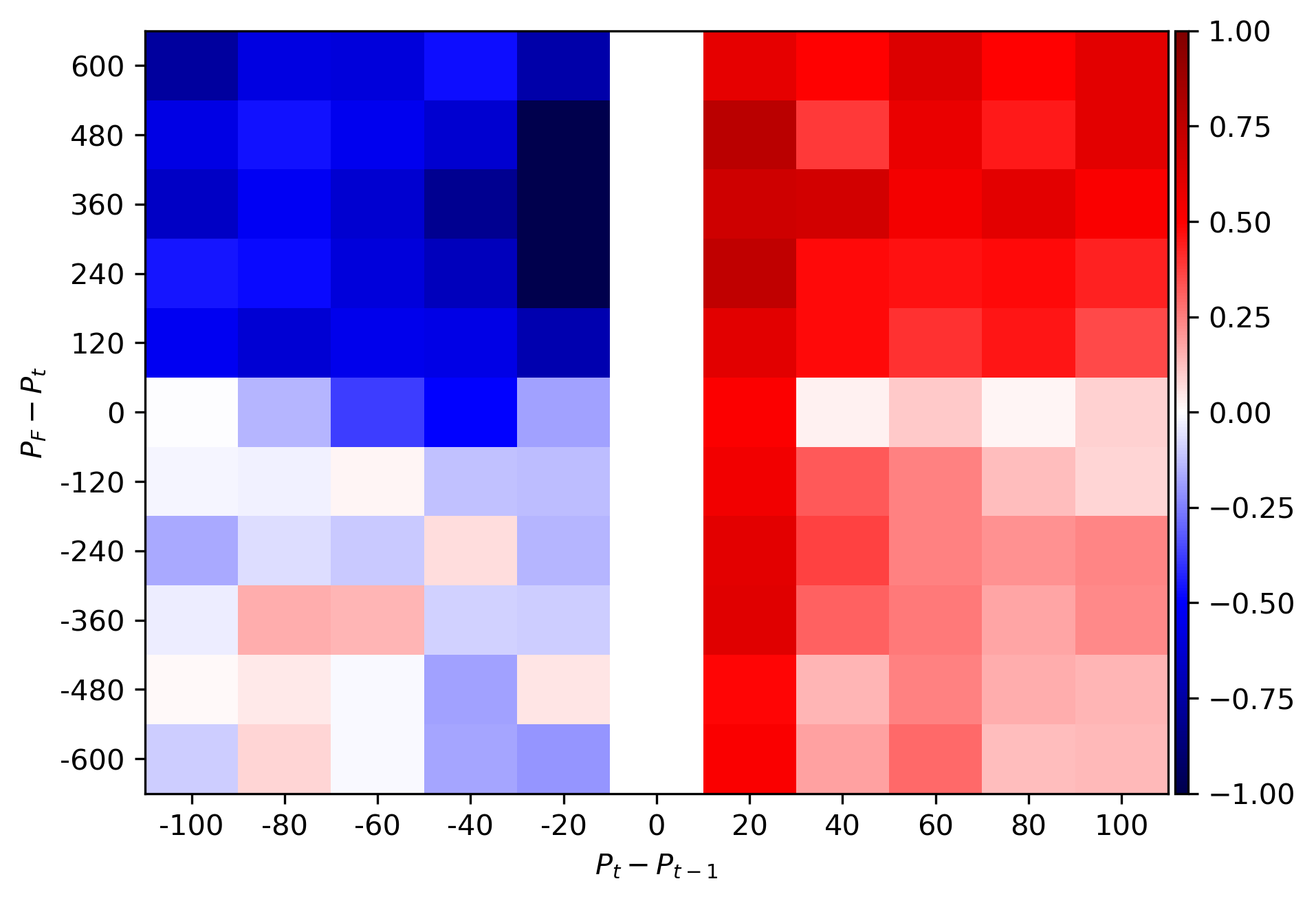}\\[1ex]
    % Riga 2
    \includegraphics[width=0.48\textwidth]{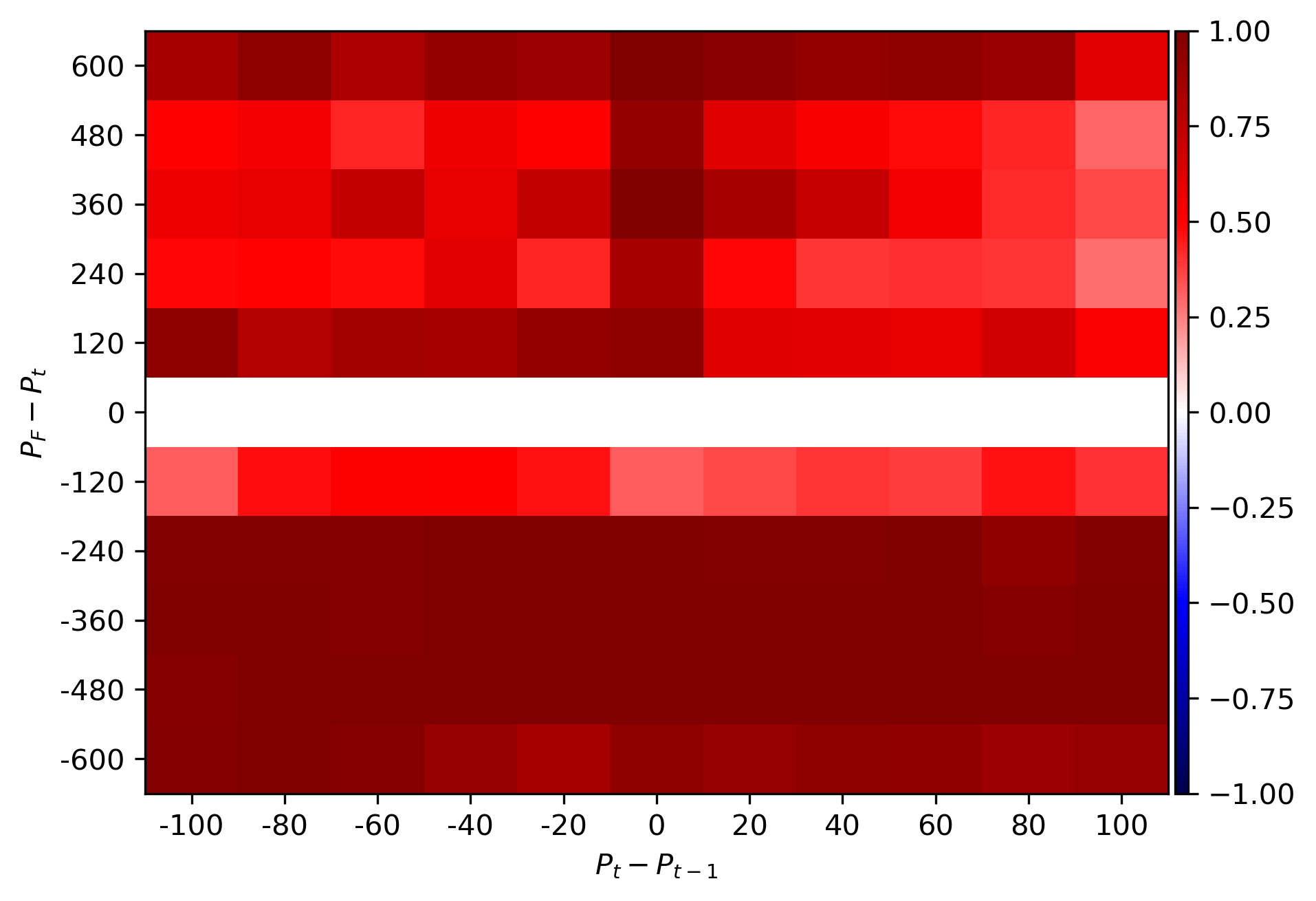}\hfill
    \includegraphics[width=0.48\textwidth]{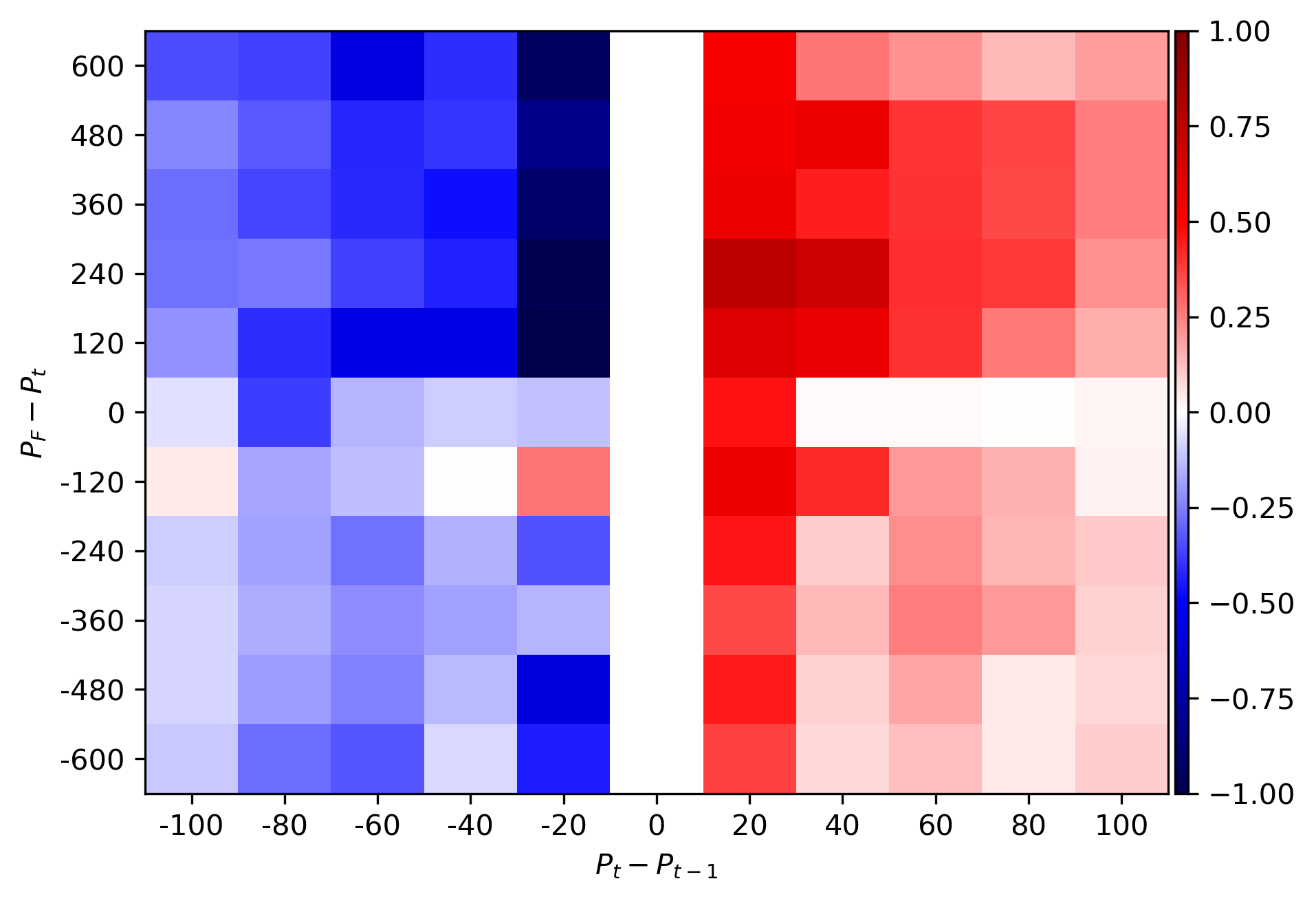}
    \includegraphics[width=0.48\textwidth]{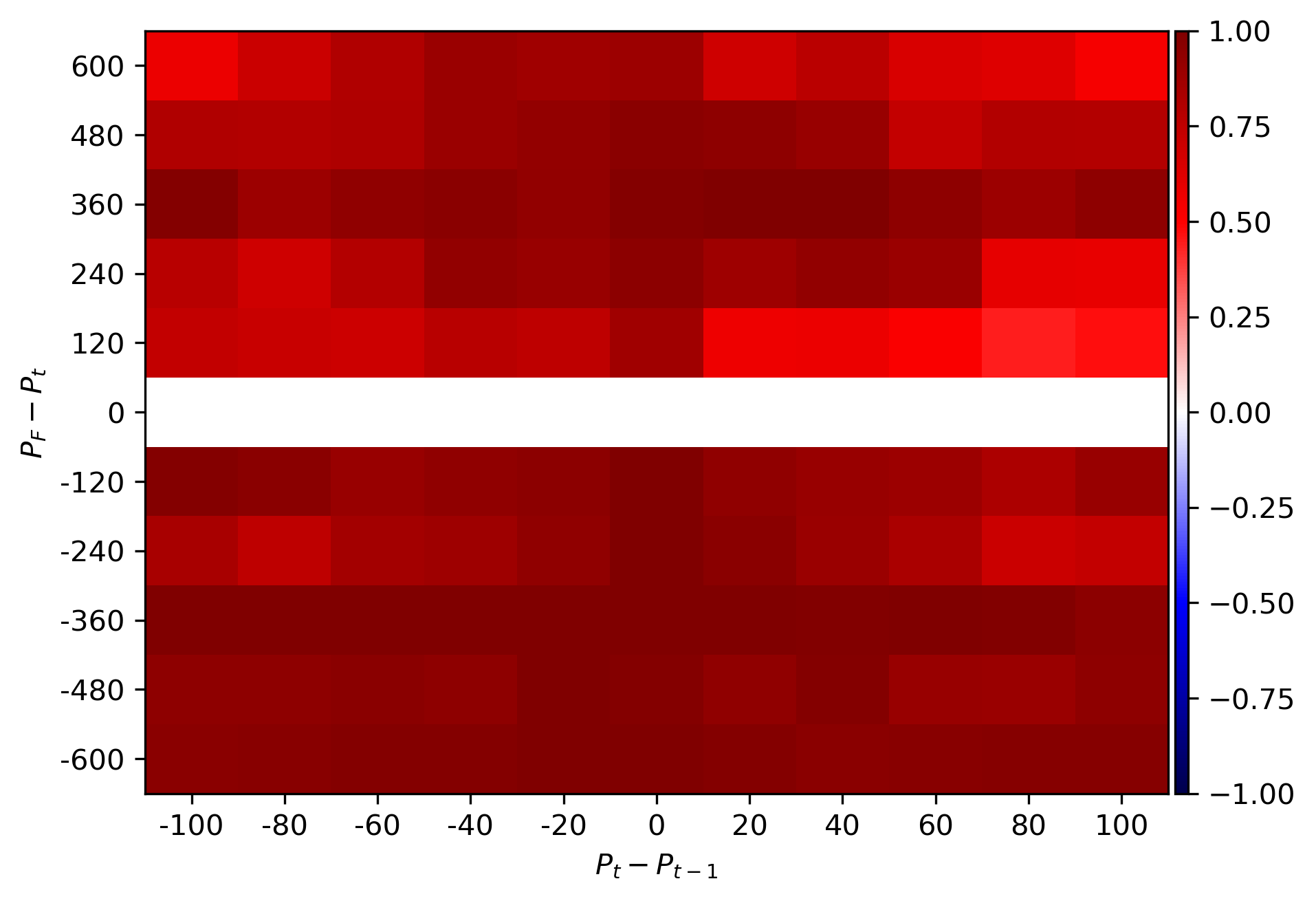}\hfill
    \includegraphics[width=0.48\textwidth]{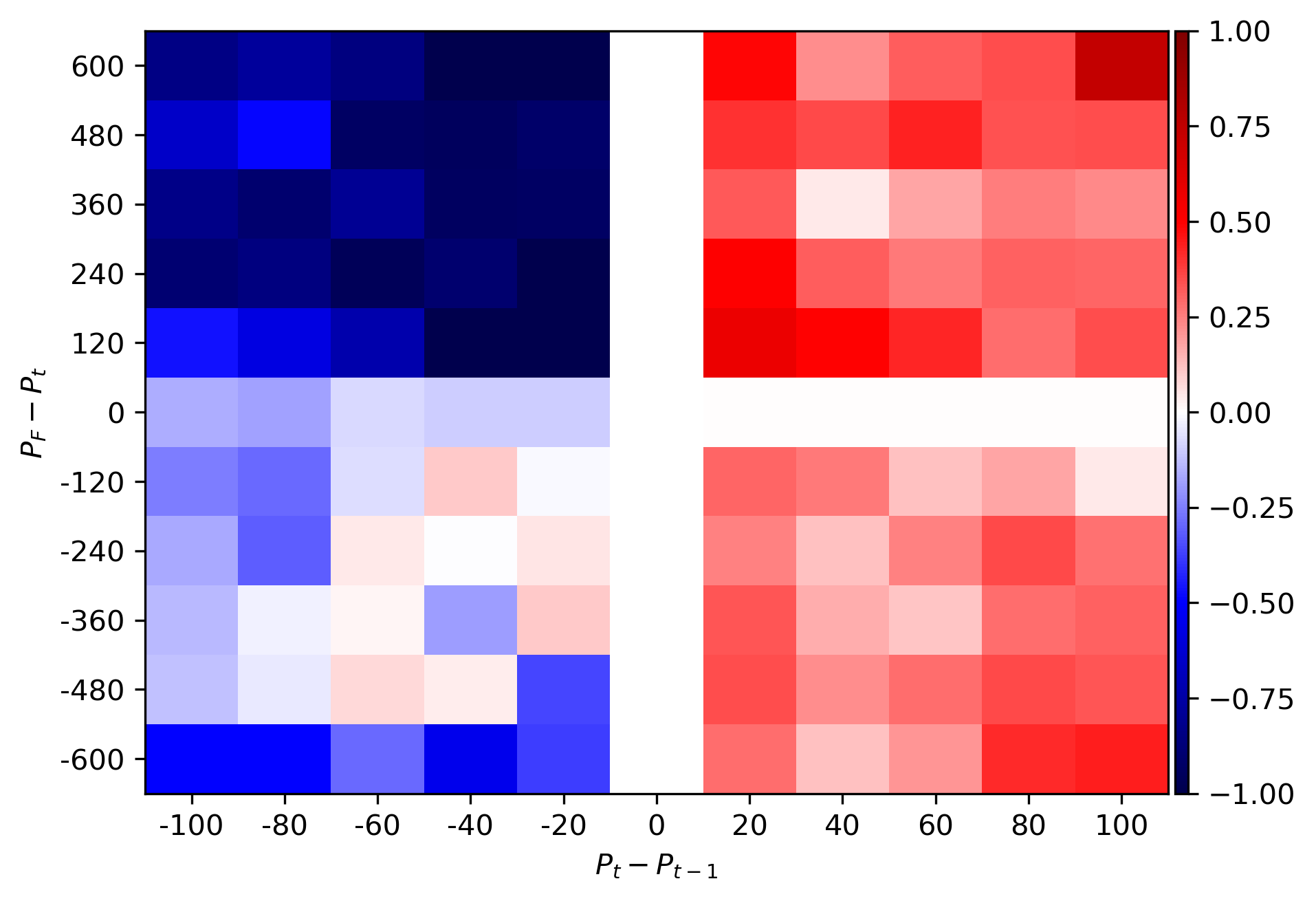}\\[1ex]
    % Riga 2
    \includegraphics[width=0.48\textwidth]{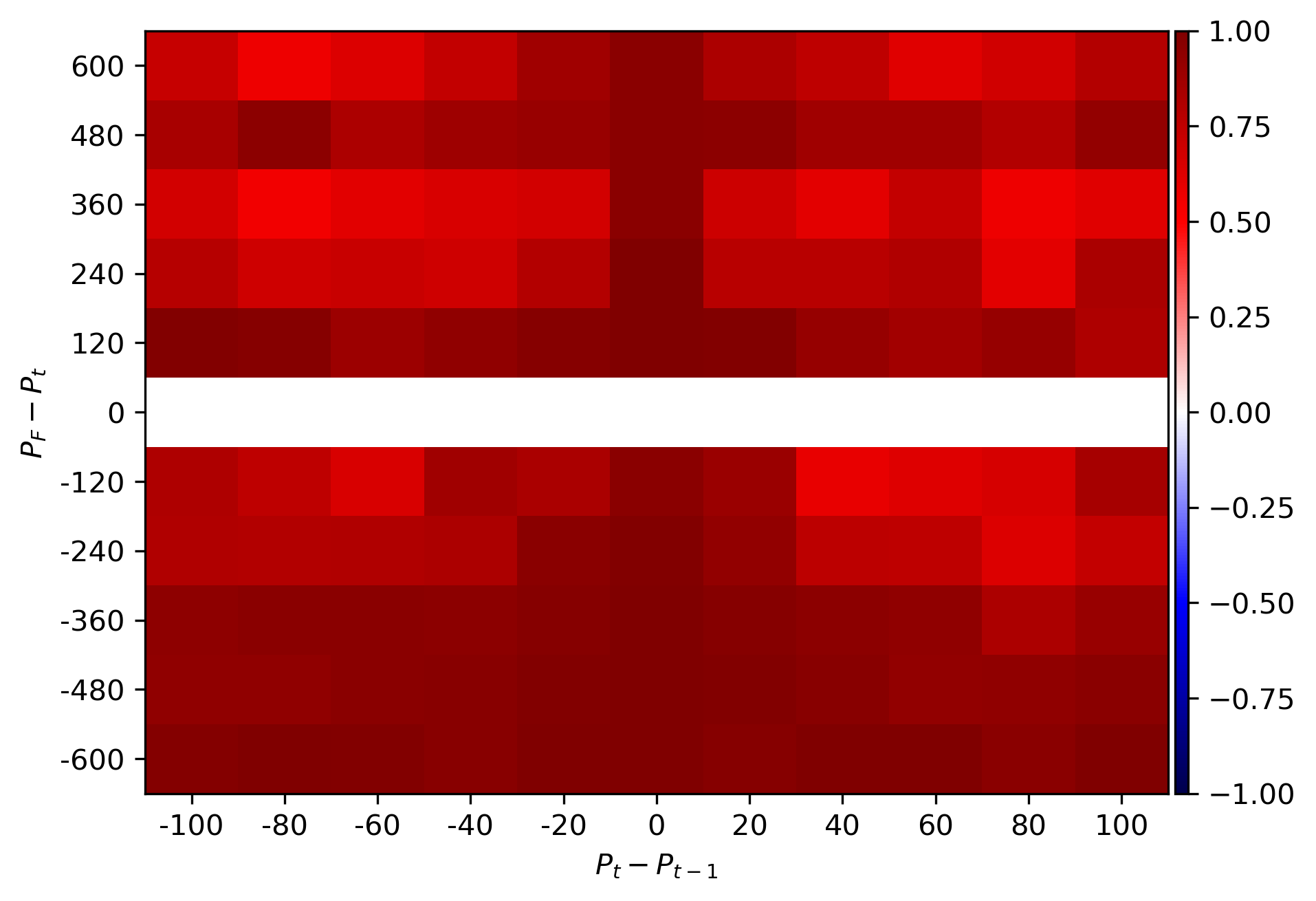}\hfill
    \includegraphics[width=0.48\textwidth]{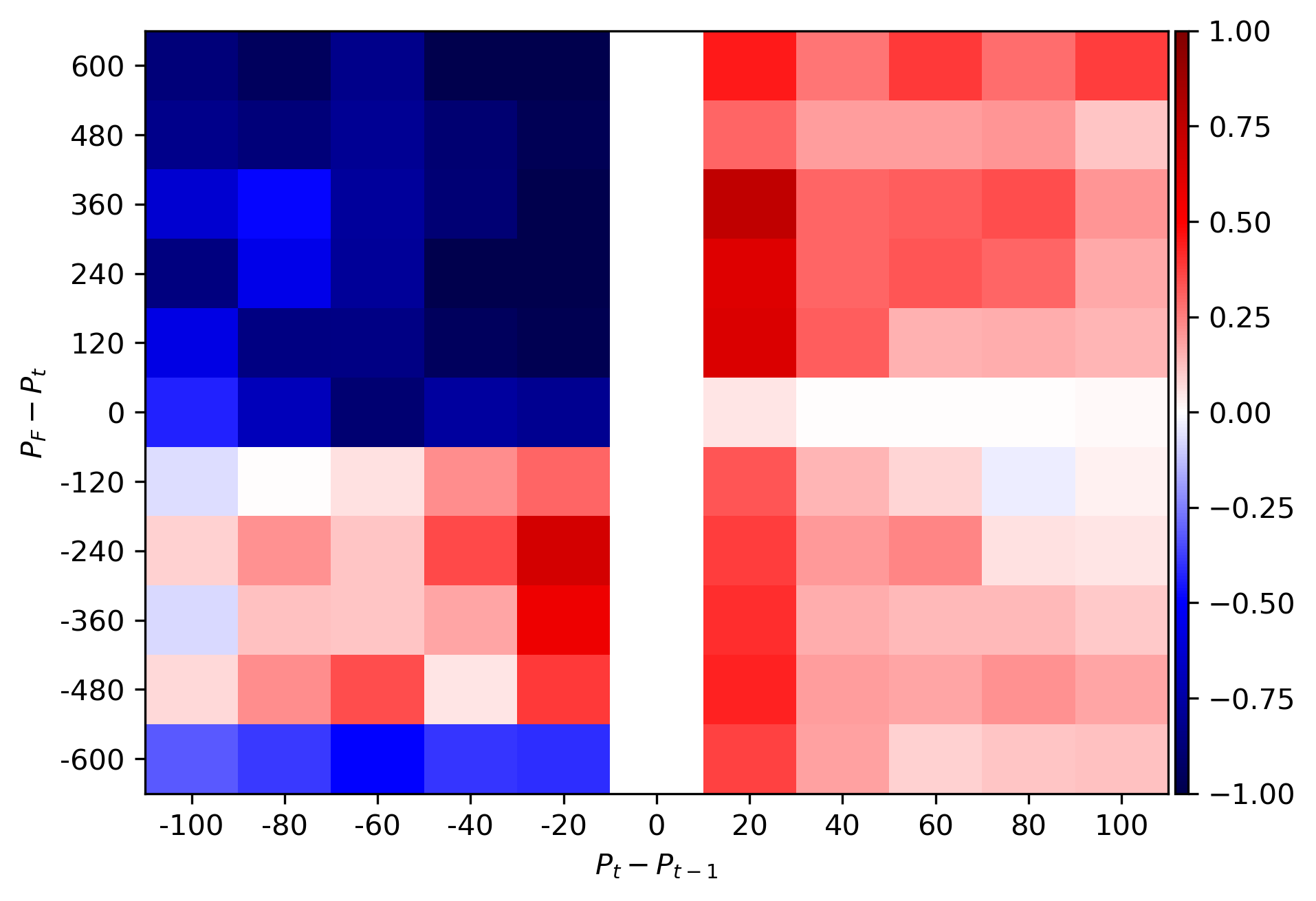}
    \caption{Left: values of $\alpha$; Right: values of $\beta$; First line: simulations for $P_t=833$; Second line: $P_t=1083$; Third line: $P_t=1583$; Fourth line: $P_t=1833$. Each subfigure reports averages over 50 repetitions with gpt-4o-mini at temperature 1.}
    \label{fig: alpha beta pt appendix}
\end{figure}

In the following, we report the results of the reaction parameters as the current price varies. As before, we present the average values of the reaction parameters $\alpha$ and $\beta$. The results for the reaction parameter $\alpha$ in equation \ref{eq: alpha}, related to fundamentalist expectations, and for the reaction parameter $\beta$ in equation \ref{eq: beta}, related to trend follower expectations, are reported in Figure \ref{fig: alpha beta pt appendix}. Consistent with previous findings, the parameter $\alpha$ always takes values between 0 and 1, while $\beta$ shows greater variability. In line with earlier results, when the price is below the fundamental ($P_t^F-P_t>0$), and the trend is negative ($P_t-P_{t-1}<0$), the trend follower acts as a contrarian, expecting a reversal. In contrast, when the trend is positive ($P_t-P_{t-1}>0$) and approaches the fundamental, $\beta$ takes values greater than zero, indicating that the trend follower expects the upward trend to continue. When the current price is above the fundamental ($P_t^F-P_t<0$), there is a stronger tendency to follow an upward trend ($P_t-P_{t-1}>0$), with $\beta>0$. When the trend is downward ($P_t-P_{t-1}<0$), both positive and negative values of $\beta$ coexist, leaving room for expectations that oppose the declining trend.

\begin{figure}[t]
    \centering
    % Riga 1
    \includegraphics[width=0.48\textwidth]{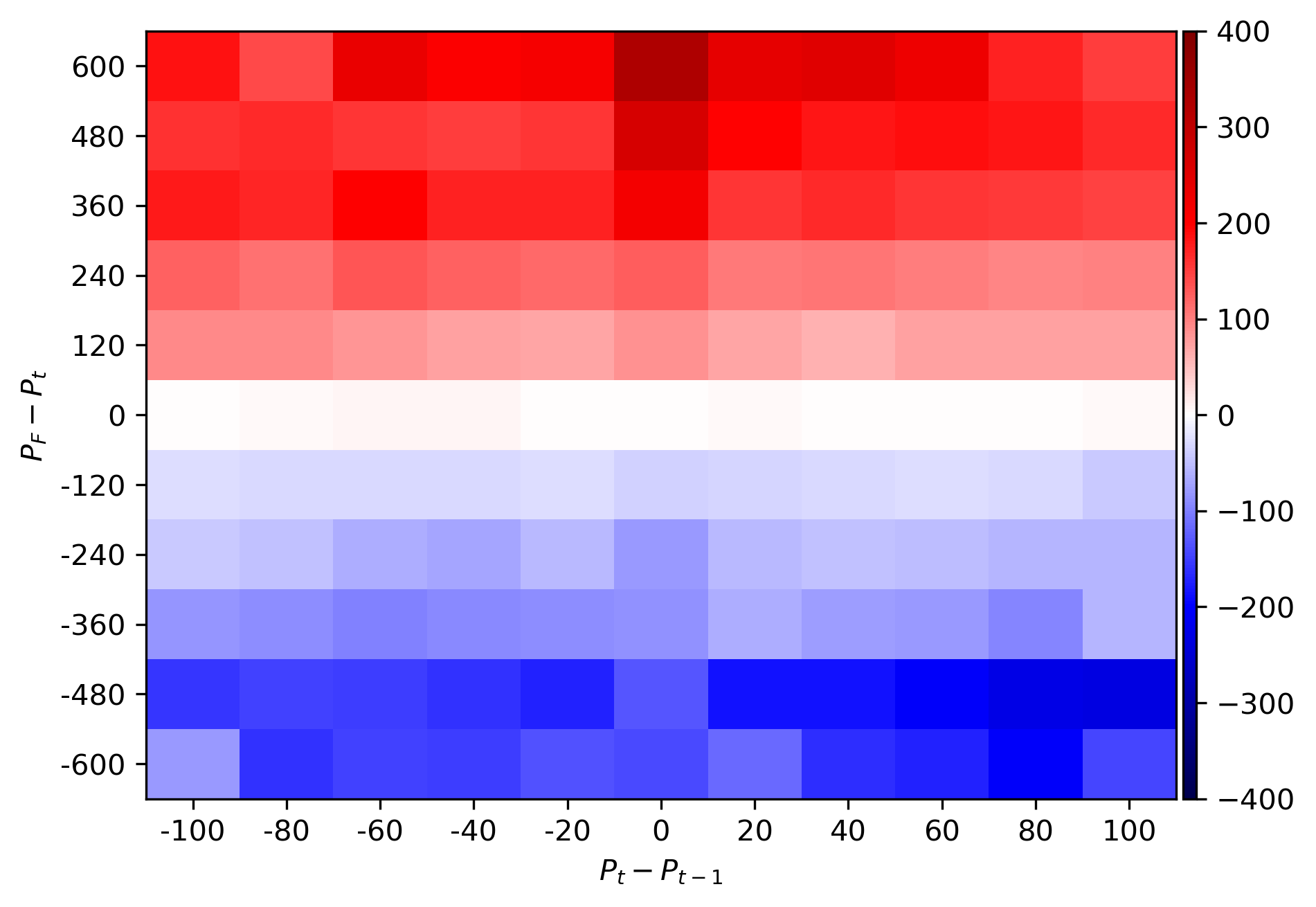}\hfill
    \includegraphics[width=0.48\textwidth]{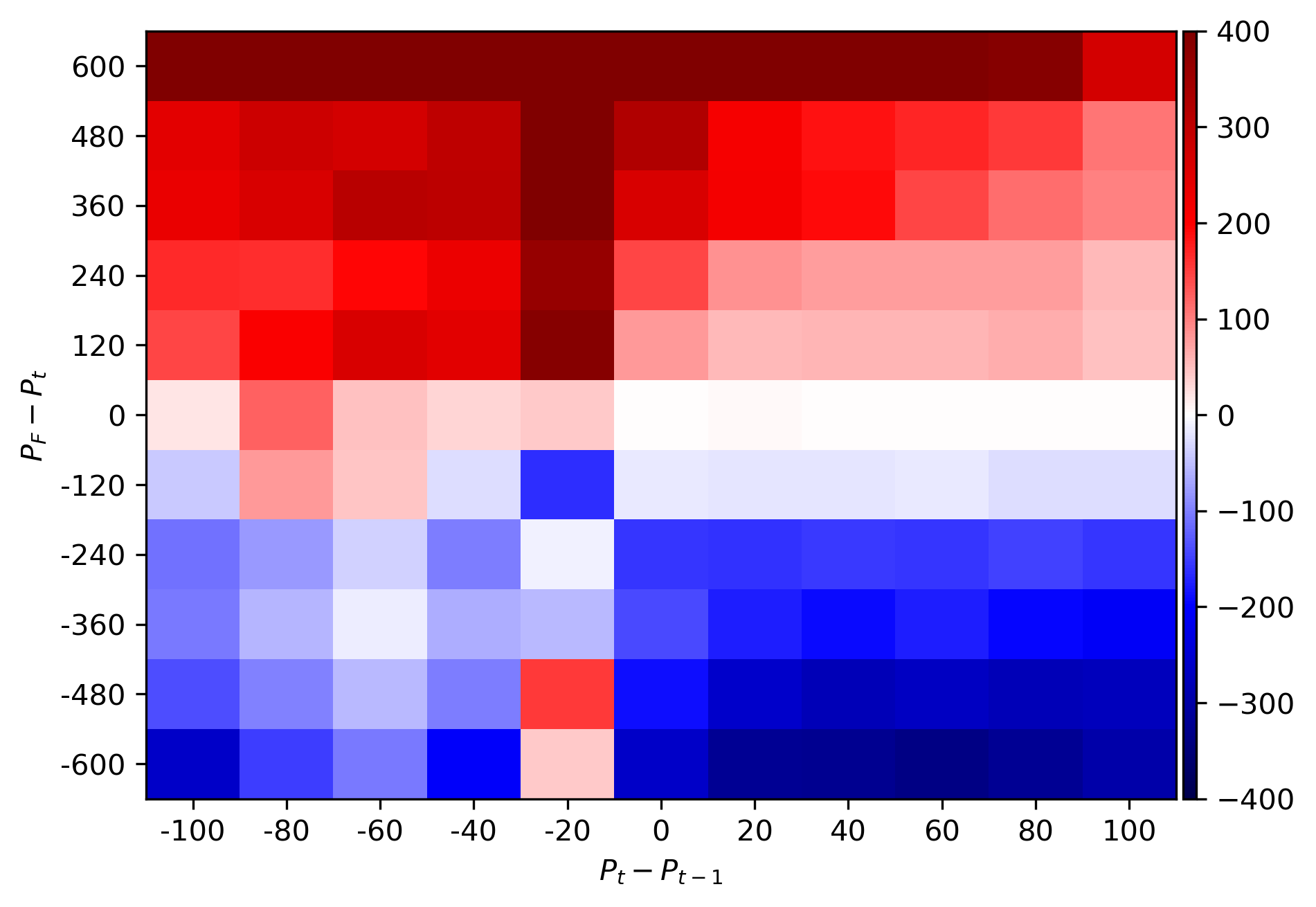}\\[1ex]
    % Riga 2
    \includegraphics[width=0.48\textwidth]{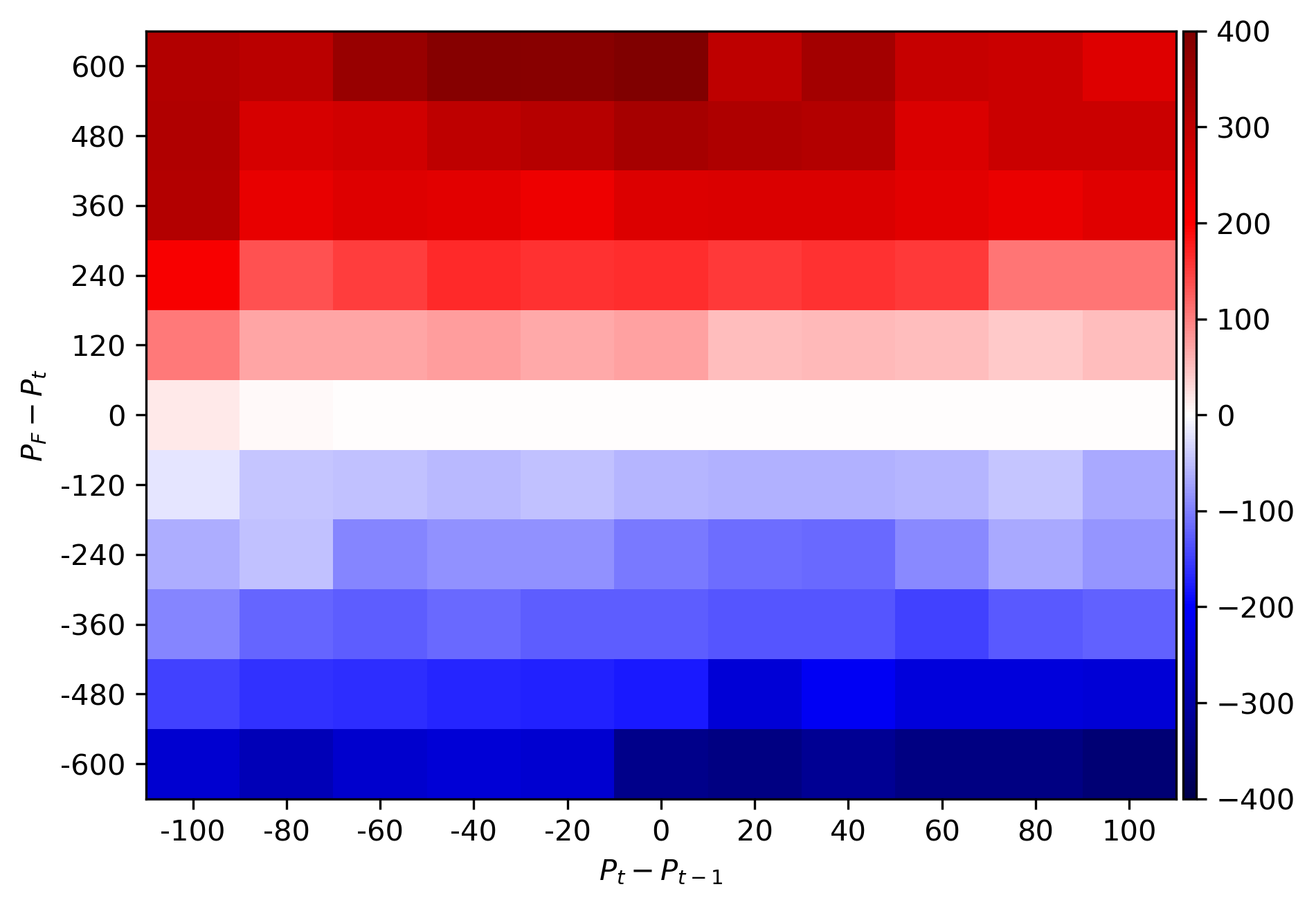}\hfill
    \includegraphics[width=0.48\textwidth]{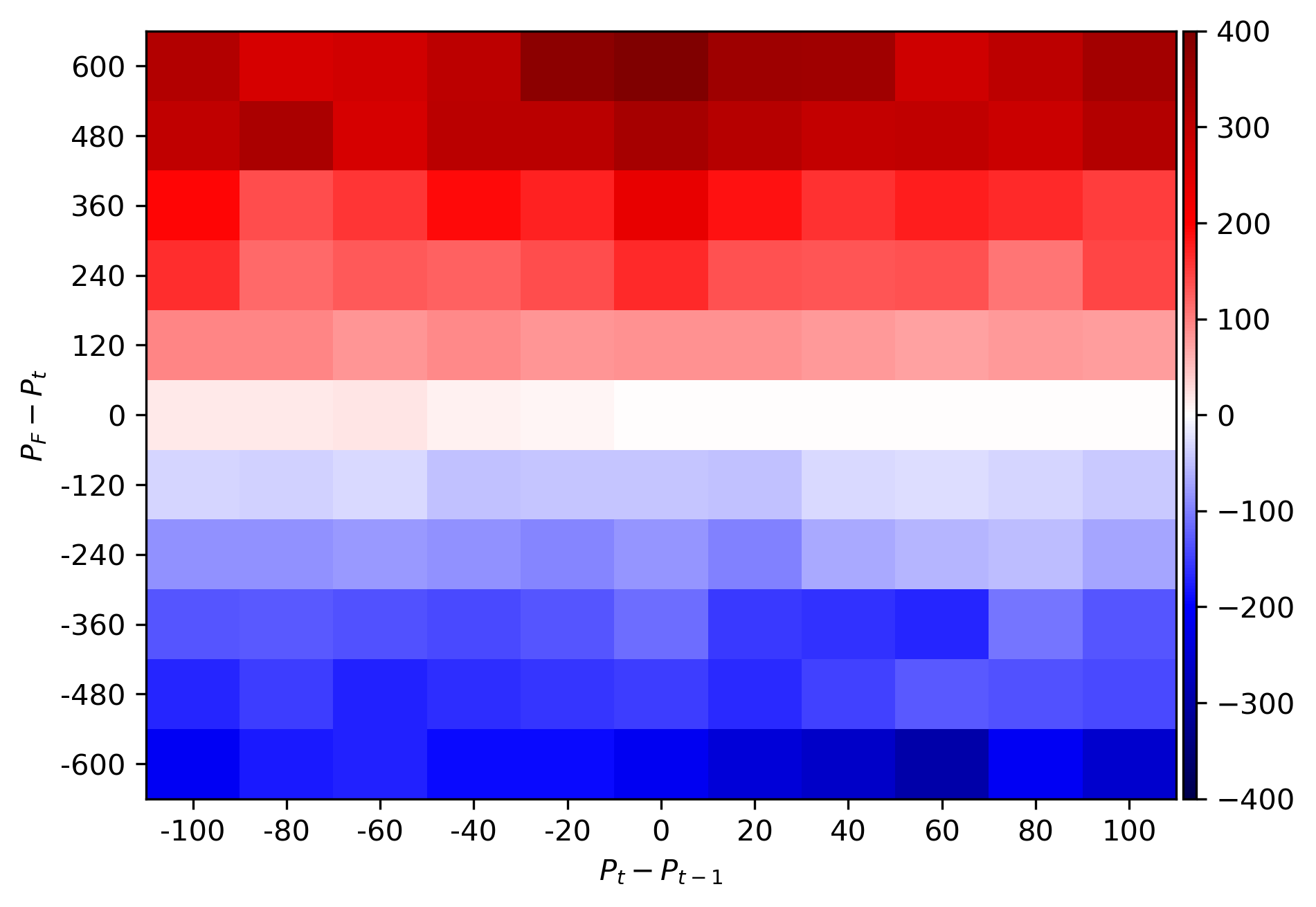}
    \caption{Expected price variation. Top-left: $P_t=833$; top-right: $P_t=1083$; bottom-left: $P_t=1583$; bottom-right: $P_t=1833$.}
    \label{fig: expected price pt appendix}
\end{figure}

Finally, we present the results for the expected price change as the reference price at time $t$ varies. Specifically, we use the weights (Figure \ref{fig: stretgie pt appendix}) and the values of the reaction parameters (Figure \ref{fig: alpha beta pt appendix}) to compute the expected price; by subtracting $P_t$ from this value, we obtain the expected price variation. The results, shown in Figure \ref{fig: expected price pt appendix}, indicate that the result remains robust to changes in the value of $P_t$.

\subsection{Dimension of simulations}\label{appendix: dimension}

\begin{figure}[H]
    \centering
    % Riga 1
    \includegraphics[width=0.48\textwidth]{figures/set_agosto_strategie_temp1_n50_pt1333.png}\hfill
    \includegraphics[width=0.48\textwidth]{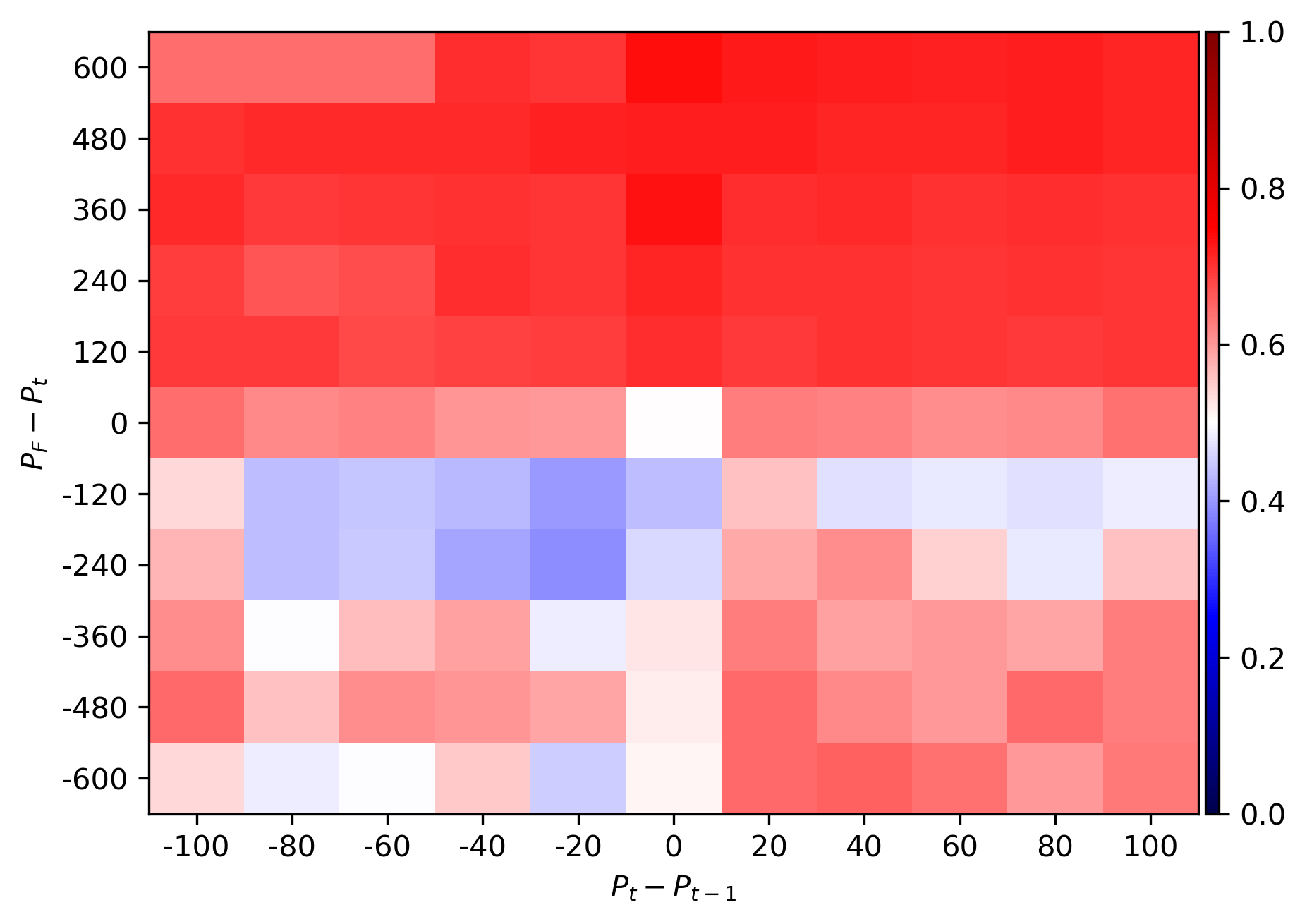}
    % Riga 2    
    \caption{Weight of the fundamentalist strategy. On the left: 50 repetitions; On the right: 100 repetitions. Each subfigure reports results with \textit{gpt-4o-mini} at temperature 1 and $P_t=1333$.}
    \label{fig: dimension}
\end{figure}

In this subsection, we examine the stability of the results with respect to the number of repetitions. In the main analysis, the results were averaged over 50 repetitions of the prompt. Here, we show the results using 100 repetitions, keeping all other parameters constant: gpt-4o-mini with temperature 1, the original prompt and $P_t=1333$. As shown in Figure \ref{fig: dimension}, the results with 50 repetitions (left) are stable and fully replicated with 100 repetitions (right).

\section{Robustness GPT parameters}
In this section, we test the robustness of the results by varying the LLM parameters specified in the Python code. Specifically, in subsection \ref{appendix: appendix 3.5} we change the model to gpt-3.5-turbo, and in subsection \ref{appendix: appendix temperature} we vary the temperature.
\subsection{GPT model}\label{appendix: appendix 3.5}

\begin{figure}[hb]
\centering
    \includegraphics[scale=0.49]{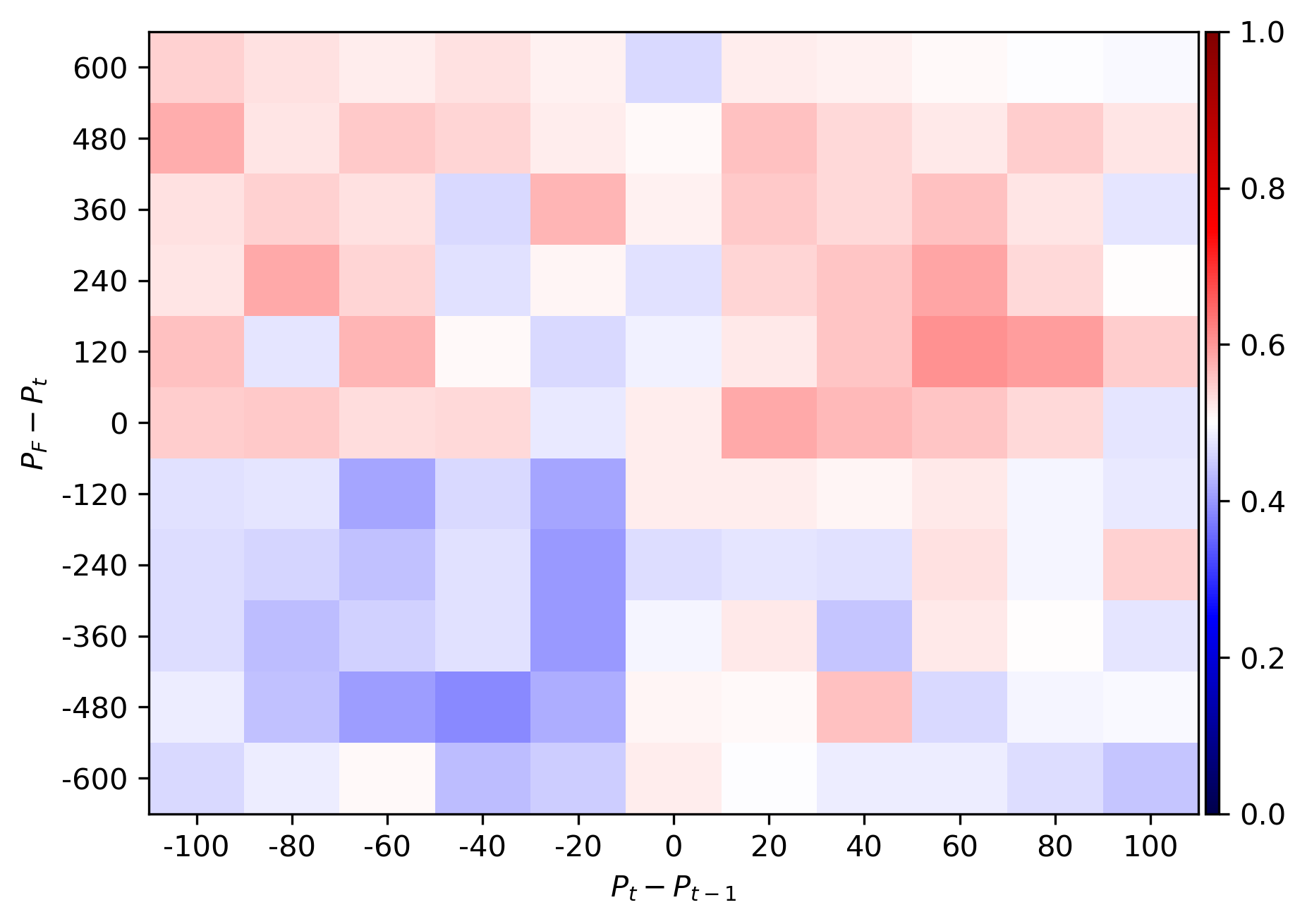}
    \caption{Weight of the fundamentalist strategy. The figure reports the averages over 50 repetitions with \textit{gpt-3.5-turbo} at temperature 1, using the original prompt and $P_t=1333$.}
    \label{fig: gpt 3.5}
\end{figure}

In this subsection, we present the main results obtained by varying the chatGPT model. We use gpt-3.5-turbo, keeping the temperature at 1, the original prompt, and $P_t=1333$. As before, results are averaged over 50 repetitions. As shown in Figure \ref{fig: gpt 3.5}, the asymmetry between the cases where the price is above or below the fundamental persists even with the model change. The absolute values are lower, reflecting greater variability in the responses across the 50 runs.

\subsection{Temperature}\label{appendix: appendix temperature}

In this subsection, we examine the robustness of the results by varying the ChatGPT temperature to 0.4 and 0.7, compared to 1 used in the main analysis. As previously mentioned, in models such as ChatGPT, the temperature parameter controls the randomness of text generation by influencing how likely the model is to select less probable tokens. At low values, close to zero, the model becomes more deterministic, consistently choosing the most probable words and producing stable, predictable, and coherent outputs, though at the expense of creativity and diversity. At higher values, typically between 0.8 and 1.2, the model explores a wider range of possible continuations, which increases variability and originality in the responses, but also increases the likelihood of inconsistencies or errors.

\begin{figure}[ht]
    \centering
    % Riga 1
    \includegraphics[width=0.49\textwidth]{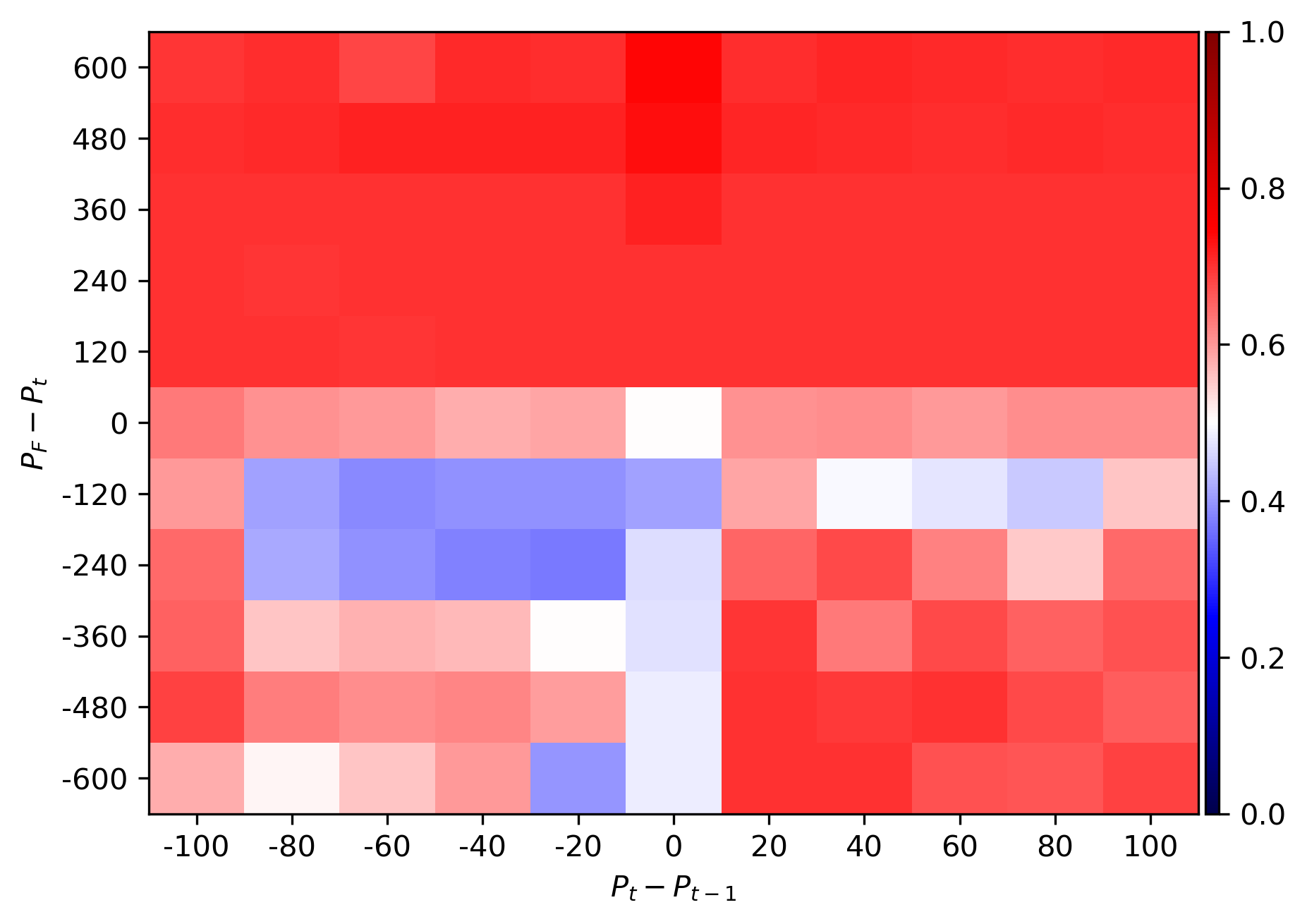}\hfill
    \includegraphics[width=0.49\textwidth]{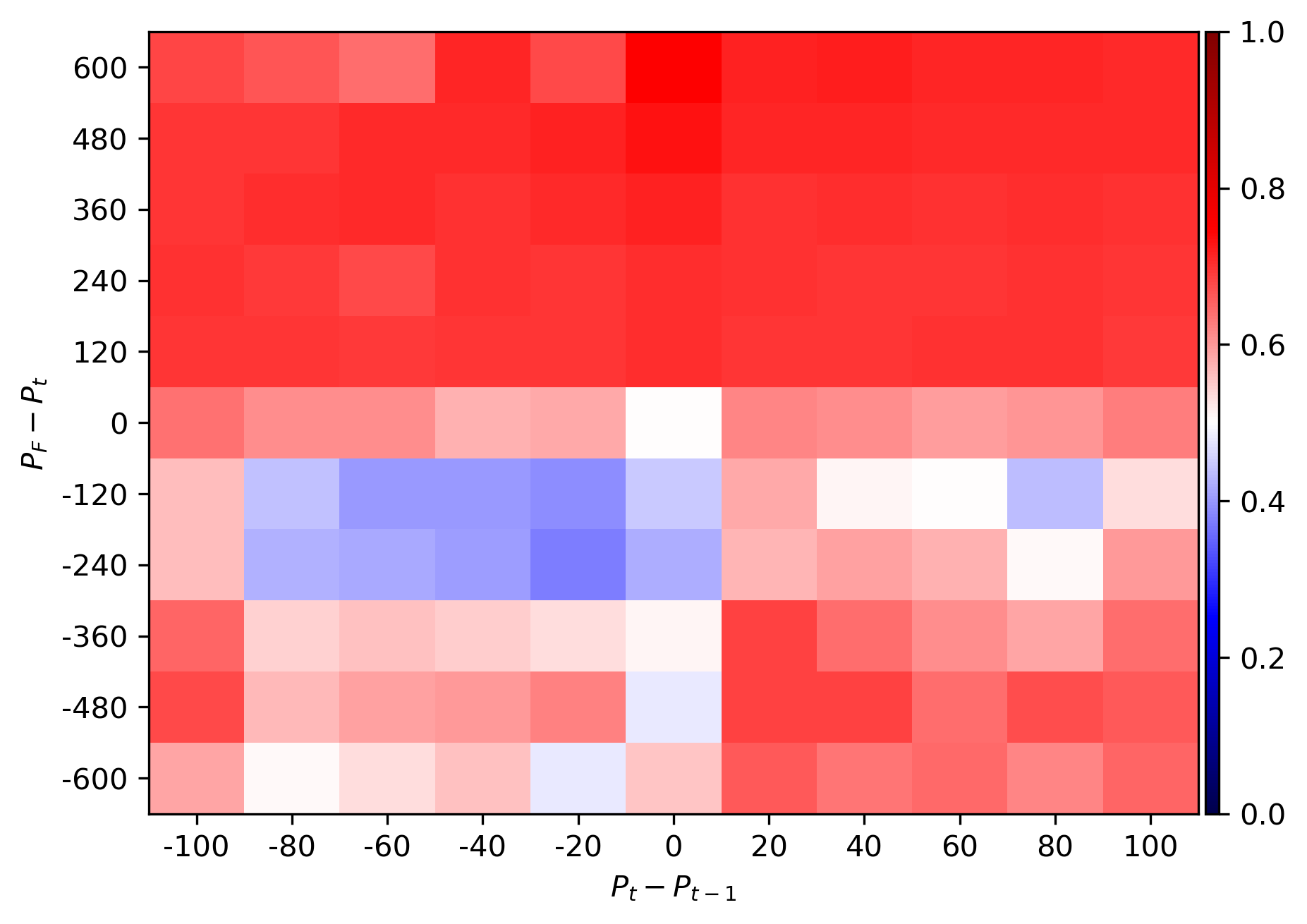}\\[1ex]
    % Riga 2
    
    \caption{Weight of the fundamentalist strategy. On the left: temperature equal to 0.4; On the right: temperature equal to 0.7.}
    \label{fig:temperature}
\end{figure}

The results are shown in Figure \ref{fig:temperature}. All other parameters are kept constant, results are averaged over 50 repetitions using gpt-4o-mini, using the original prompt, and with $P_t=1333$. For both temperatures -0.4 on the left and 0.7 on the right - the main results are confirmed: the asymmetry between cases where the price is above or below the fundamental is evident, with a predominance of fundamentalists when the price is above.

\end{document}